\documentclass[12pt,twoside, a4paper]{article}

\usepackage[english]{babel}
\usepackage{graphicx,epstopdf,epsfig}
\usepackage{amsfonts,epsfig,fancyhdr,graphics,amsmath,amssymb}

\usepackage[unicode]{hyperref}

\newcommand{\tluste}[1]{\mbox{\mathversion{bold}$ #1 $}}

\newcommand{\X}[0]{{\tluste{x}}}
\newcommand{\Y}[0]{{\tluste{y}}}

\newcommand{\ol}[1]{\mbox{$\overline{{#1}}$}} 
\newcommand{\ul}[1]{\mbox{$\underline{{#1}}$}}

\newtheorem{theorem}{Theorem}

\newtheorem{definition}[theorem]{Definition}

\begin{document}



\begin{center}
\textbf{\Large Contribution of Interval Linear Algebra to the Ongoing Discussions on Multiple Breath Washout Test}\\
\vspace{1em}
Jaroslav Hor\'{a}\v{c}ek\footnote{Department of Applied Mathematics,
	Charles University, Prague, Czech Republic
	(\url{horacek@kam.mff.cuni.cz}). Supported by Charles University Grant Agency (GAUK) grant no. 174815.}, V\'{a}clav Kouck\'{y}\footnote{Department of Paediatrics, 2nd Faculty of Medicine, Charles University, Prague, Czech Republic (\url{vaclav.koucky@seznam.cz}). Supported by Charles University Grant Agency (GAUK) grant no. 174815.} and Milan Hlad\'{i}k \footnote{Department of Applied Mathematics,
	Charles University, Prague, Czech Republic (\url{hladik@kam.mff.cuni.cz}). Supported by GA\v{C}R grant no. P403-18-04735S}
\end{center}

\begin{abstract}
In the paper the interval least squares approach to estimate/fit  data with interval uncertainties is introduced. The solution of this problem is discussed from the perspective of interval linear algebra. Using the interval linear algebra carefully, it is possible to significantly speed up the computation in specialized cases. The interval least squares approach is then applied to lung function testing method -
Multiple breath washout test (MBW). It is used for algebraic handling of uncertainties arising during the measurement. Surprisingly, it sheds new light on various aspects of this procedure -- it shows that the precision of currently used sensors does not allow verified prediction. Moreover, it proved the most commonly used curve to model the nitrogen washout process from lung to be wrong. Such insight contributes to the ongoing discussions on the possibility to predict clinically relevant indices (e.g., LCI).
\end{abstract}


\textbf{AMS.} 65G40, 62J05, 92C50  



\section{Introduction 1 -- Multiple Breath Washout test (MBW)} \label{intro-sec}
First works concerning multiple breath washout test date back to '40s and '50s \cite{cotes2009lung}.
In those days the method faced crucial limitations which prevented its use in clinical practice. The precision of sensors was not sufficient to measure low gas concentrations accurately and also the computational power of digital computers was insufficient to handle problems described using too many parameters (much of mathematical work was still done manually). With increasing power of sensors and computers MBW was reborn in '90s.

MBW is a very promising lung function test since it does not require any specific breath maneuvers. The only requirement is the regular tidal breathing with no leaks. 
This makes it applicable to the wide age range of patients including
infants, who undergo this test in sleep (either artificial or natural).

In contrast to the conventional methods (e.g, spirometry, bodypletysmography), MBW is able to evaluate even the most peripheral airway disease.  Its high sensitivity to the most peripheral airway changes has been shown in most of chronic lung diseases (e.g. bronchial asthma, cystic
fibrosis, primary cilliary dyskinesia etc.) \cite{davies2008}, \cite{green2011}, \cite{macleod2009}.

\begin{figure}[h]
	\centering
	\includegraphics[width=6cm]{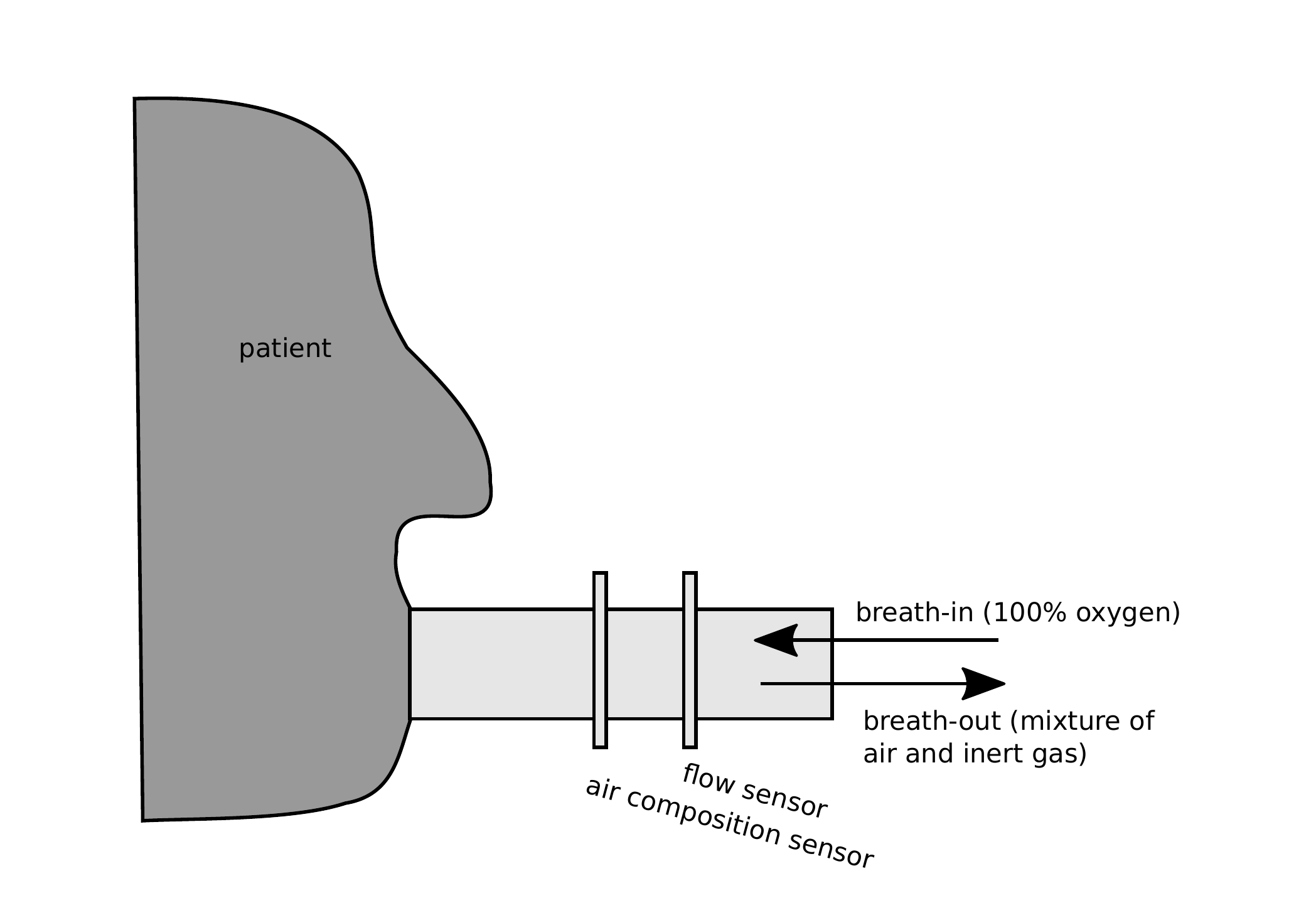}
	\caption{\footnotesize Schematic depiction of the washout phase.}
	\label{fig:exhalyzer}
\end{figure}

The test consists of two phases -- washin and washout. During the first
phase lung is filled with an inert gas (sulphur hexafluoride $SF_6$,
helium $He$ or resident inert gas -- nitrogen $N_2$), during the second phase the inert gas is being
washed out by air or by 100\% oxygen (depending on the inert
gas used). Concentration of the respective inert gas, volume
of exhaled gas and flow are measured in real time. 
The measurement is stopped after reaching a certain level of inert gas concentration within lung (usually 2.5\% of the initial gas concentration). 
The pattern
of inert gas concentration decrease gives information about
the homogeneity of ventilation and thus about the patency
of the most peripheral airways.  The washout procedure can be seen in Figure \ref{fig:exhalyzer}. 

In our work we focus on use of the nitrogen multiple breath washout test ($N_2$-MBW). 
Although, the $SF_6$ has been used for much longer time, the use of nitrogen as inert gas has many advantageous properties:  
\begin{itemize}
	\item $SF_6$ is not used in medicine, so it must be specially prefabricated, $N_2$ is naturally present in the surrounding air 
	\item $SF_6$ is an exogenous inert gas, which needs to be washed in to the lungs
	\item $SF_6$ is not routinely available in medical settings, it is quite expensive and have a severe green-house effect  
\end{itemize}
Contrarily, $N_2$ is naturally present in the
surrounding air and in lung (so called endogenous inert gas) -- there is no need for
washin phase. Moreover, $N_2$ is present also in poorly ventilated areas of lung, which makes the evaluation of ventilation inhomogeneity in severely affected patients more accurate.
A small drawback is that the nitrogen is not ideal because of its solubility in blood and its back-diffusion from tissues and blood to the lung during washout phase. 

\begin{figure}[h]
	\centering
	\includegraphics[width=10cm]{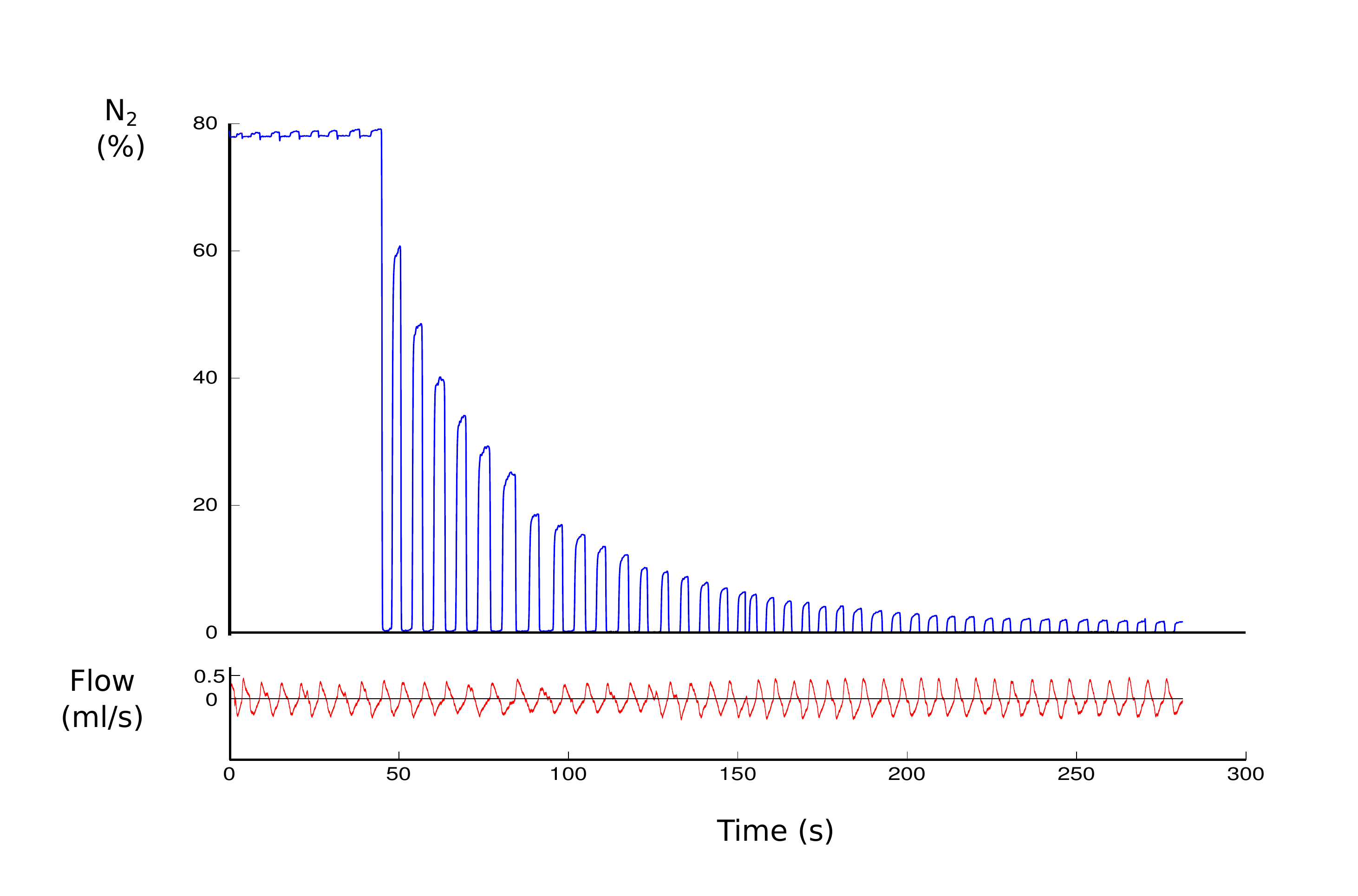}
	\caption{\footnotesize Nitrogen concentration (top curve) and air flow (bottom curve) in time measured during the nitrogen washout process. In about 50th second the washout phase begins (the first large drop of $N_2$ concentration).}
	\label{fig:nitrogenflow}
\end{figure}

The main output is depicted in Figure \ref{fig:nitrogenflow}. These are two graphs -- actual flow (bottom curve) and decreasing nitrogen concentration (top curve) measured in each \emph{time slice}. These data are further used for computing clinically significant indices (FRC, LCI, Scond, Sacin, etc.). Some of them will be mentioned further in the text.  
The advantage of MBW is its high sensitivity to the  early stages of various lung diseases. That enables early therapeutic intervention. 
 There are studies that describe a typical evolution of the mentioned indices for a given disease \cite{gustafsson2007peripheral}.

\section{LCI and FRC}

Functional residual capacity (FRC, defined as volume of air in lung at the end of tidal exhalation) and lung clearance index (LCI $n$ defined as number of lung volume turnovers required to wash out the inert gas to $n\%$ of its initial concentration) are 
currently the most important indices derived from MBW data. If we omit the correction for deadspace ventilation, the FRC is calculated as follows:
$$ FRC = \frac{V_{N_2 out}}{C_{N_2 start} - C_{N_2 end}}, $$
where $V_{N_2 out}$ is the total volume of expired $N_2$, $C_{N_2 start}, C_{N_2 end}$ are concentrations of nitrogen in the start and end point of FRC computation respectively. The index LCI is calculated simply as

\begin{equation}
LCI = \frac{V_{out}}{FRC}, 
\label{lci}
\end{equation}
where $V_{out}$ is the total volume of expired air. 
It is necessary to specify how we define the terminal breath (the end of measurement). 
 It is defined as the first of three consecutive breaths with end tidal concentration of inert gas under a preset level; historically it is 2.5\%.
The corresponding LCI index is then marked LCI2.5.
The whole washout process up to 2.5\% might be too time consuming. Making it difficult for uncooperative patients to finish the MBW test properly. 

The FRC relates to the size of lung. The LCI states how many air volumes (equal to FRC) exchanges are necessary to clean the lung from the inert gas (more specifically to reach the level of 2.5\% of initial inert gas concentration). The LCI index seems to be very useful to evaluate the homogeneity of lung ventilation (the most peripheral airways included).

Currently, there are some ongoing  discussions about possibility to use of the level 5\% as the end of washout (LCI5). 
We also would like to contribute to the discussion with our results. 



\section{Introduction 2 -- Interval methods}
The history of interval analysis is actualy quite similar to MBW. It was developed in '50s but it took some time (until '90s) before it started to be used practically (mainly because of insufficient computational power of machines). 

The basic notion is an \emph{interval} (denoted in boldface), which is for our purpose a real closed interval containing all real numbers within the lower and upper bounds,
$$\tluste{x} = [\ul{x}, \ol{x}] = \{ x, \ \ul{x} \leq x \leq \ol{x}\}.$$
The interval works in a verified way. That means it contains some desired value (e.g., a physical constant, a root of polynomial) for sure, however, it is not known where exactly the value lies. With intervals, the arithmetic can be defined as follows:

\noindent Let us have two intervals $ \X = [ \ul{x}, \ol{x} ] $ and $ \Y = [ \ul{y}, \ol{y}] $ then
\begin{align*}
\X + \Y & = [ \ul{x} + \ul{y}, \ol{x} + \ol{y} ],\\ 
\X - \Y & = [ \ul{x} - \ol{y}, \ol{x} - \ul{y} ],\\
\X \, * \, \Y  & = [ \min(S), \max(S) ], \quad \textrm{where} \  S = \{  \ul{xy}, \, \ul{x} \ol{y}, \, \ol{x} \ul{y}, \, \ol{xy} \},\\ 
\X \ / \ \Y & = \X * (1 / \Y), \quad \textrm{where} \  1/\Y = [1/\ol{y}, 1/\ul{y}], 
0 \notin \Y. 
\end{align*}
Interval arithmetics can be (carefully) incorporated in our computational problems (solving systems of equations, integration, data fitting) in order to preserve the verified nature of our data. 
 
\begin{figure}[h]
	\centering
	\includegraphics[width=10cm]{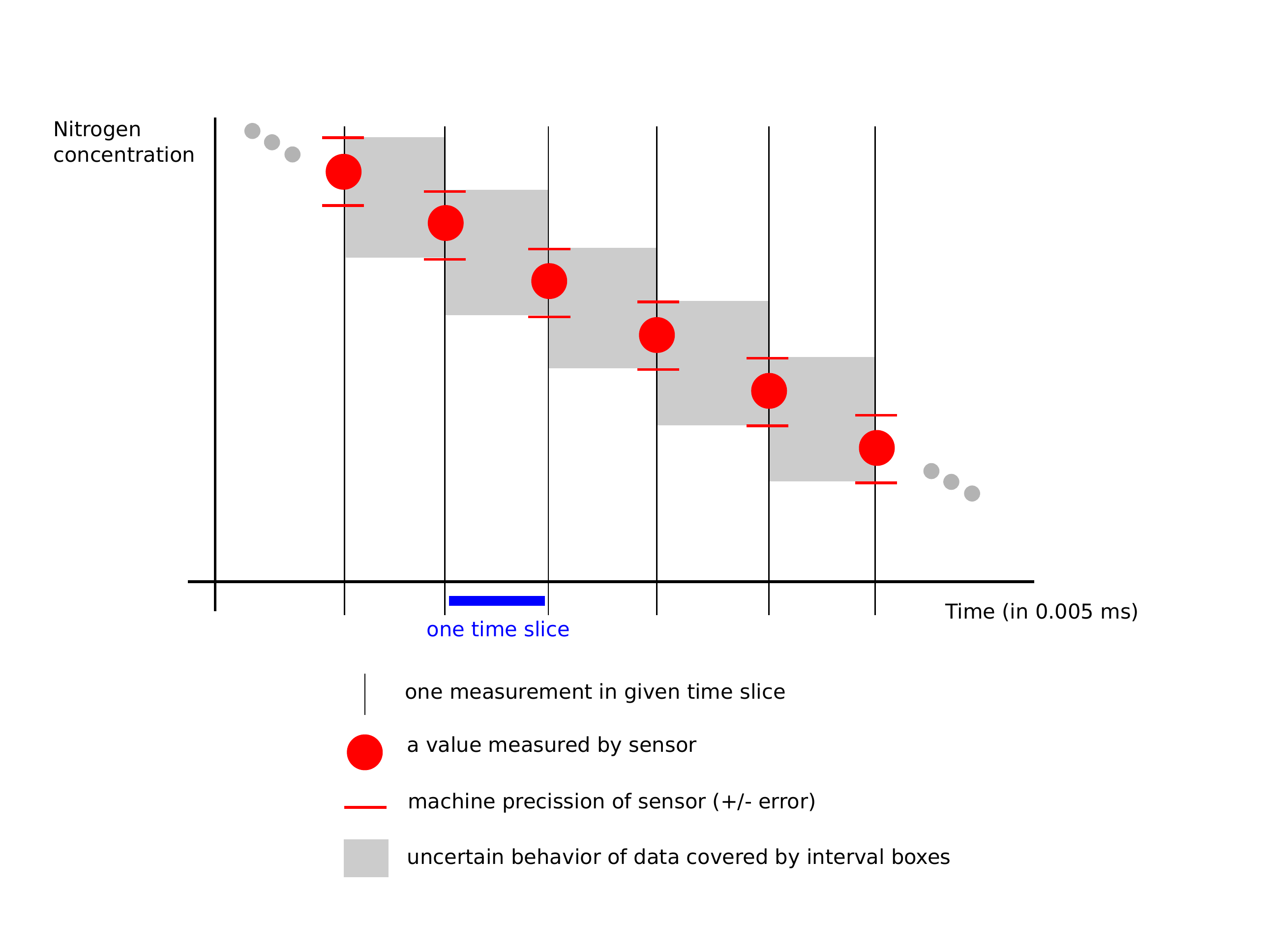}
	\caption{\footnotesize An illustration of how  intervals occur when having discrete sampling and given measurement error (a segment of a nitrogen washout curve).}
	\label{fig:intervaldata}
\end{figure}

 Interval computation is used when dealing with measurement and rounding errors. In the MBW procedure there are many sources of such errors: 
 
	\begin{itemize}
		\item Imprecision of sensors 
		\item Changing viscosity and humidity of air
		\item Time shift of signals
		\item Interaction with deadspace air
		\item Physiological noise (heart pulse, hick-ups, leaks)
		\item Irregular breathing pattern, apnea
		\item Computer and machine rounding errors
		\item etc.
	\end{itemize} 
Unknown distributions and interplay of the  mentioned uncertain variables will result in intervals with unknown distribution. Hence they will only provide verified lower and upper bounds.  The situation is depicted in figure \ref{fig:intervaldata}. 

Regarding data estimation one can usually think of some form of regression. Of course, the meaning of "regression on interval data" needs to be specified first. We will provide the definition in Section \ref{regression}. We will see that interval analysis provides an interesting tool for dealing with such uncertainties algebraically (using means of interval linear algebra).

More about interval analysis and its use can be found in e.g., \cite{JauKie2001, kearfott1996interval, moore:introduction}.

\subsection{Nitrogen concentration in peaks}

To obtain interval bounds for nitrogen concentration in peaks (end of breaths) we must first locate the end of breaths. For that purpose we developed our own algorithm \cite{Breathends} which is able to outperform the existing 
state-of-the-art approaches and even commercial software (Spiroware). After localization of the breath ends the imprecision of machine sensors must be incorporated.   
We used Exhalyzer D machine by Ecomedics, Duernten, Switzerland, that does not measure nitrogen concentration directly. It computes the nitrogen concentration (in \%) according to the formula \cite{jensen2013standard}:

$$100 =  N_2\%  + O_2\% + CO_2\% + Ar\%,$$ 
where $Ar\% = N_2\% \times  0.0093/0.7881$ and
where the concentrations of nitrogen, oxygen, carbon dioxide and argon in inspired and expired air are supposed to sum up to 100 \%. With argon concentration fixed. That together gives

$$ N_2\% = \frac{1}{1.0118} (100 - O_2\% - CO_2\%), $$
where all parameters are in percents. 

According to the manufacturer, the $O_2$ sensor has accuracy 0.3\% and the $CO_2$ sensor has accuracy 5\%. 
From that we can derive a verified interval bounding the nitrogen concentration in each time slice  $\tluste{n_i}$
according to the formulas

\begin{equation}
\ul{n}_i  = \frac{1}{1.0118}(100 - 1.003* O_2\% - 1.05*CO_2\%), \label{ninterval1} 
\end{equation}
\begin{equation}
\ol{n}_i  = \frac{1}{1.0118}(100 - 0.997* O_2\% - 0.95*CO_2\%). \label{ninterval2} 
\end{equation}
From $100\%$ the minimal possible concentrations of $O_2, CO_2$ were subtracted to obtain an upper bound on nitrogen concentration and the maximal values were subtracted to obtain a lower bound.

\section{Regression on interval data}
\label{regression}
Various authors approached the topic of regression on interval data, e.g, \cite{cerny2013}, \cite{de2004new}, \cite{hladik2012}, \cite{tanaka1998interval}. Behind the interval regression or interval estimation the following general definition can be seen.

\begin{definition}
	A result of the  multi-linear interval regression on (interval) data tuples $$(\X^i_1, \X^i_2, \ldots, \X^i_n, \tluste{y^i})$$ is generally $$\tluste{r}(x_1, x_2, \ldots, x_n) = \tluste{p}_1 x_1 + \tluste{p_2} x_2 + \dots + \tluste{p_n}x_n,$$ where
	$\tluste{p} = (\tluste{p}_1, \ldots, \tluste{p}_n)^T$ are interval parameters. 
\end{definition}  

\begin{figure}[h]
	\centering
	\includegraphics[width=10cm]{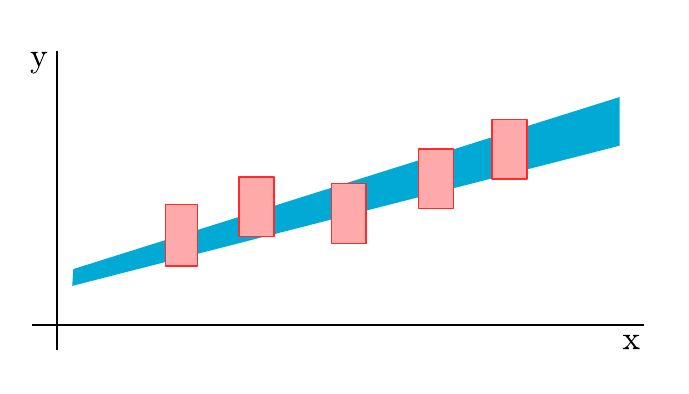}
	\caption{\footnotesize An example of $\tluste{r} = \tluste{p}_1 x + \tluste{p}_2$. The band actually forms an interval line, which passes through the interval boxes.}
	\label{fig:regression}
\end{figure}

The resulting $\tluste{r}$ can be viewed as a multi-dimensional band. A two dimensional example can be seen in Figure \ref{fig:regression}.

As it was explained, there are various types of interval regression. They vary in computation of interval parameters $\tluste{p}$. For example, $\tluste{p}$ could be computed in such a way to force the band $\tluste{r}$ to contain all the data tuples, or at least to cross all the interval data. For our purpose the interval least squares approach is the most meaningful. 

\begin{definition}
	For a given data: an $m \times n$ interval matrix $\tluste{X}$, where its $i$-th row is the tuple $$ (\X^i_1, \X^i_2, \ldots, \X^i_n),$$ and an $m$-dimensional column vector $\tluste{y}$, where its coefficients are $\tluste{y}^i$, the interval parameters $\tluste{p}$ of the interval least squares estimation are defined in the following way,
	$$\tluste{p} = \Box \{ p : X^TX p = X^Ty \ \textrm{for some} \ X \in \tluste{X}, y \in \tluste{y}  \},$$
	where $\Box \{ \cdot\}$ is the tightest possible enclosure of a given set by an $n$-dimensional box (interval vector).
\end{definition}

\subsection{Computation of interval least squares and our improvement}
When we are given real data $(X, y)$, the least squares parameters $p$ can be obtained by solving $ X^TX p = X^T y$. However, this is not the recommended approach since the condition number of the matrix of the new system is squared. There are various possibilities how to approach this problem (QR-decomposition, Krylov subspace methods). They both rely on orthogonality, however when we are tackling the general interval data $(\tluste{X}, \tluste{y})$ the orthogonality of two interval vectors makes no sense, hence these methods are of no use.
When we try to apply the first mentioned approach to the interval case $ \tluste{X}^T\tluste{X} p = \tluste{X}^T \tluste{y}$ it seems nice, since there are a lot of methods for solving interval linear systems \cite{Hla2014b}, \cite{horacek:oils}, \cite{moore:introduction}, \cite{neumaier1990interval}. 
Unfortunately, multiplication of two interval matrices results not only in quadratic condition number but also in exceptional growth of interval widths, therefore the obtained solution $\tluste{p}$ would generally be useless.  
The state of the art approach is mentioned in e.g., \cite{neumaier1986linear} it is based on solving the following system
\begin{equation}
\left( \begin{array}{cc}
I & \tluste{X} \\
\tluste{X}^\top & 0 \\
\end{array} \right)
\left( \begin{array}{c}
p\\
p_2\\ 
\end{array} \right)
=
\left( \begin{array}{c}
\tluste{y}\\
0\\ 
\end{array} \right). 
\label{state-of}
\end{equation}
The enclosure of parameter vector $\tluste{p}$ appears as the first $n$ components of the obtained enclosure.  
From $X$ we form much larger square matrix, that is why, we call it supersquare or supsquare approach. It can be seen that much larger system of interval equation needs to be solved.

\begin{figure}[h]
	\centering
	\includegraphics[width=8cm]{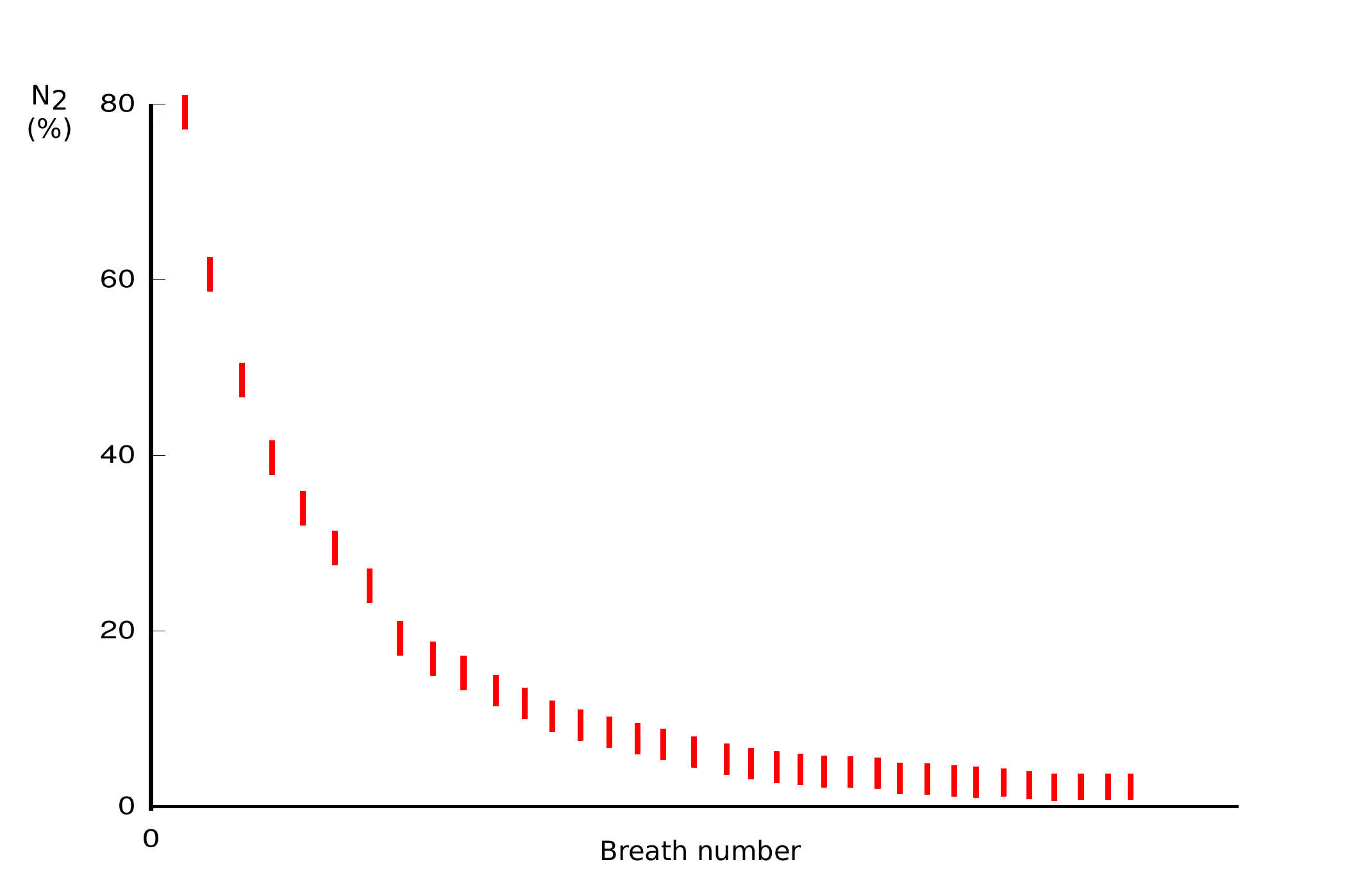}
	\caption{\footnotesize Ilustration of decreasing concentration of nitrogen in peaks bounded with intervals. From such data the $\tluste{X}, \tluste{y}$ for the regression will be formed.}
	\label{fig:exponential}
\end{figure}

Later, we want to use regression with nonlinear models that are linearizable, therefore the data $(\tluste{X}, \tluste{y})$ formed out of the MBW data depicted in Figure \ref{fig:exponential} will have a certain shape:
		\begin{itemize}
			\item $X=\tluste{X}$ is thin 
			\item Intervals are to be found only in the right-hand side $\tluste{y}$ 
			\item $X^\top X$ is going to be small $n \times n$, $(n=2, 3, 4)$, depending on the model used (see the Table \ref{tab:models} in advance) 
			\item Depending on the linearization used, $X$ might consist of integers only (ones, numbers of breaths or its powers)
			\item $X, \tluste{y} > 0$ (component-wise)			
		\end{itemize}

All these are really favorable properties. That is why we asked whether it is possible to design a method returning tighter enclosures than (\ref{state-of}).  Unfortunately, we were not able to find such method. 
We believe that it is a really hard task since the mentioned properties are also in favor of (\ref{state-of}).
However, we were able to rewrite the formulas to obtain algorithms that are much faster.

\subsection{Case $2 \times 2$} When $X$ is an integer matrix of size $m \times 2$, then $X^TX$ is of size $2 \times 2$. We can apply the state-of-the-art supsquares approach, however, in this case the "not-recommended" approach of solving the interval normal system of equations $X^TX p = X^T \tluste{y}$  approach may pay off. This actually means computing some verified enclosure of $\tluste{p}$, where
$$ \tluste{p} = (X^T X)^{-1} X^T \tluste{b}.$$ 
When computing the inverse matrix, fractions can occur and therefore possibly machine non-representable numbers can occur. That is why we also need to compute in a verified way with intervals. Nevertheless, it is advantageous to postpone the interval computation as far as possible, because the classical arithmetic is usually faster (e.g., in Octave or Matlab). In this case we use the simple shape of the $2\times 2$ matrix inverse
$$ (X^T X)^{-1} = \left( \begin{array}{cc} 
a & b \\
c & d \\
\end{array} \right)^{-1} = \frac{1}{ad - bc} \left( \begin{array}{cc} 
d & -b \\
-c & a \\
\end{array} \right).$$  
It is possible to compute $X^T X$ in floating point arithmetic since $X$ contains only integers, and similarly for $ad -bc$.

When we compute the expression $(X^T X)^{-1} X^T \tluste{y}$ we multiply $\tluste{y}$ with an interval matrix $X^T$, that unfortunately causes large growth of interval radii. And then we multiply it again with the matrix $(X^T X)^{-1} $ which causes another growth. Much more suitable way is to compute the whole expression as
$$ \left((X^T X)^{-1} X^T \right) \tluste{y}, $$ 
that is multiplying the matrices first and then multiplying with $\tluste{b}$.
In conclusion, the enclosure of $\tluste{p}$ can be computed as 
$$ (M X^T) (\tluste{q} \tluste{y} ) \supseteq \tluste{p},$$
where $$ M = \left( \begin{array}{cc} 
d & -b \\
-c & a \\
\end{array} \right) , \ \tluste{q} = \Box\left(\frac{1}{ad - bc}\right),$$
the symbol $\Box(\cdot)$ stands for tightest enclosure of the given expression by an interval with machine representable bounds.

The system  (\ref{state-of})  can be solved by any cited method computing enclosures of interval systems.  We used the method within the Octave interval package \cite{heimlich2016gnu}. 
We tested the difference between the two mentioned approaches on random systems for sizes $m=50, 100, 150, 200, 250$, which represent the ceiling for the maximum number of breaths generally occurring during MBW testing. For the purpose ot the testing, $X$ consisted of two columns, the first consisted of numbers 1 to $n$ and the second of ones. For the right hand side $\tluste{y}$ we first generated random intervals with fixed radius 1 and then these intervals were placed along a random line.  Both methods were tested for each size on 100 systems. To show the difference, our method was tested with and without postponing of interval operations. In all cases methods computed enclosures for $\tluste{p}$ of the same width. However,  computational times were different, they are displayed in Table \ref{tab:case2x2}. In the table a difference between postponed and non-postponed interval computation can be clearly seen. 

\begin{table}[h]
	\renewcommand\arraystretch{1.3} 
	\small
	\centering
	\tabcolsep=1em
	\begin{tabular}{cccc}
		\hline
		m & \texttt{supsq} & \texttt{normal} & \texttt{postponed}\\[0.4ex]
		\hline
		50 &  0.036  & 0.030 & 0.0072\\
		100 & 0.062 & 0.031 & 0.0073\\
		150 & 0.11  & 0.032 & 0.0075\\
		200 & 0.18 & 0.033 & 0.0077\\
		250 & 0.28 & 0.034 & 0.0078\\
		\hline
	\end{tabular}
	\caption{\footnotesize Computation times (in seconds) for $2\times2$ systems for supsquares approach (\texttt{supsq}) and solving interval normal equations without (\texttt{normal}) and with postponing the interval operations (\texttt{postponed}).}
	\label{tab:case2x2}
\end{table}

\subsection{Case $3 \times 3$ and larger} It would be more complicated to find similar formula for an inverse of a general square matrix. That is why this time we refrain from postponing interval computations and enclose $X$ directly with tight intervals (e.g., with radii $10^{-15}$). We again compare it with the supsquare approach. The obtained enclosures of $\tluste{p}$ are again the same and the time computations are displayed in Table \ref{tab:case3x3}. This method is still faster than the supsquares approach.     

\begin{table}[h]
	\renewcommand\arraystretch{1.3} 
	\small
	\centering
	\tabcolsep=1em
	\begin{tabular}{ccc}
		\hline
		m & \texttt{supsq} & \texttt{normal}\\[0.4ex]
		\hline
		50 &  0.037 & 0.032 \\
		100 & 0.064 & 0.032 \\
		150 & 0.11 & 0.033\\
		200 & 0.18 & 0.034 \\
		250 & 0.28 & 0.035 \\
		\hline
	\end{tabular}
	\caption{\footnotesize Computation times (in seconds) for $3 \times 3$ systems for supsquares approach (\texttt{supsq}) and solving interval normal equations (\texttt{normal}).}
	\label{tab:case3x3}
\end{table}


\section{Our data}
We collected the data from real patients measured for medical purposes. The measurement technique adhered to ERS/ATS recommendation and the standard operation procedure for $N_2$-MBW \cite{jensen2013standard}, \cite{robinson2012consensus}.

The three necessary conditions to obtain reliable data were: 
\begin{itemize}
	\item Patients have sufficiently regular breathing pattern during measurement
	\item There is no leakage during the measurement
	\item Wash-out phase is finished (nitrogen is washed out to a preset level -- 2.5\%) 
\end{itemize}

The data was captured using Exhalyzer D machine by Ecomedics, Duernten, Switzerland.

We included 15 raw data files (A-files) from healthy volunteers (50\% of males ) with mean age 12.4 years. Additionally, we included 12 A-files from patients with cystic fibrosis (40\%  of males) with mean age 10.6 years.
The study was approved by the institutional ethical committee of University Hospital Motol, Prague. The legal representatives of patients gave written informed consent. In all A-files breath ends were detected using our own algorithm \cite{Breathends}.
Corresponding end tidal nitrogen concentrations were expressed as intervals according to the formulas (\ref{ninterval1}) and (\ref{ninterval2}). The pre-washout parts of the data were automatically trimmed.

\section{Questions we asked}
After long discussions we stated a few questions that are interesting from both clinical and mathematical point of view. The important and still discussed question is the behaviour of the nitrogen washout curve in time. There is an observable difference between the healthy and diseased persons, however the objective description is still missing. The long duration of washout (especially in severely affected patients) limits the feasibility of the test especially in small children (toddlers and pre-schoolers). Currently, the premature cessation of the washout (before reaching 2.5\% of the starting nitrogen concentration) prevents us from analysing the data. The possibility to derivate some substitute indices computable from incomplete washout curve would be of great benefit.


\section{In search for a model}

One of the main goals is to determine the shape of the nitrogen washout curve. In another words, we try to derive the following function
$$ f(n), \ \textrm{for} \ n=1,2,\ldots$$
where $n$ is the number of a peak (the initial peak has number 1) and the function $f$ returns a nitrogen concentration in each peak $n$ (it can be an interval concentration). Such function we call nitrogen \emph{washout curve model}. This goal was addressed earlier in \cite{rossing1967mathematical} using a simplified model of lungs. They were not able to compute models with more parameters due to the limited computational power (they handled many calculations manually). Their approach could be described as "bottom-up". A similar approach but for a different goal can be seen e.g., in \cite{tawhai2001multibreath}.   

Our approach is slightly different, we could call it "top-down". Using a computer we explore the most frequent mathematical models of decay and try to fit the existing medical data with them. From the fitting it will be hopefully possible to obtain more information about the real behaviour of the nitrogen washout process and such knowledge will help to better predict the behaviour of the incomplete measurement.   

Of course, there can be some outliers in the data (e.g., false breaths) that could prevent a "perfect" fit. We can use the \emph{iterative refinement} procedure -- first, the data are fitted, then the worst outlier is discarded and the data are fitted again. We tried such refinement with 1 up to 5 iterations. 
We discuss the practical use of the refinement later. 

\subsection{Center data}
In the previous sections we showed how to derive verified interval data from our measured real patient data. We applied this procedure for all datasets. First, to have at least rough idea of the washout curve model, classical least squares data fitting was applied on centered data (real data obtained when instead of each interval its center is taken). We were interested in fitting curves for which the process of good fitting can be transformed to solving a linear system of equations. When we are talking about a quality of fit we need to measure it somehow. The typically used measure is \emph{mean square error} (MSE), which measures the mean of squared distances from model fit to real data. More specifically, we use rMSE  which is the square root of MSE. We fit the data in least squares manner. 
MASE is another measure of quality of fit that we use. It measures the quality of fit of a model in contrast to the naive predictor (a function that predicts for the next step the same value that just occurred in the current step).

If we evaluate the measurements visually, we could detect "exponential"-like decay in all data. An example could be seen in Figure \ref{fig:exponential}. Many papers and books (also the medical software shipped with the machine Exhalyzer D) describe this decay with an exponential function \cite{david2003clinical}. This is one of the classical fitting models. When talking about classical fitting models we tried to find the one most suitable among them.

From the large collection of models \cite{curvehandbook} we selected the following model candidates fulfilling the visual criteria first. They are written in Table \ref{tab:models}. In the left column there is a shortcut by which we address a model, in the second column the mathematical description and in the third column the parameters that need to be computed to fit a given dataset. As already mentioned, all of these models can be linearized. For a detailed description of this process for each model see \cite{curvehandbook}.

\begin{table}[h]
	\renewcommand\arraystretch{1.3} 
	\small
	\centering
	\tabcolsep=1em
	\begin{tabular}{ccc}
		\hline
		model& function $f(x)$ & parameters\\[0.4ex]
		\hline
		{\tt exp} &  $ae^{(bx)}$ & $a, b$ \\
		{\tt explin} & $a + bx + ce^x$ & $a, b, c$\\
		{\tt pow} & $ax^b$ & $a, b$ \\
		{\tt exppow} & $ a x^b  c^x$& $a, b, c$\\
		{\tt log} & $a + b\log(x)$ & $a, b$\\
		{\tt loglin} & $a + bx + c\log(x)$ & $a, b, c$\\
		{\tt explin} & $a + bx + ce^x$ & $a, b, c$ \\
		{\tt explog} & $a + b\log(x) + c e^{x}$ & $a, b, c$ \\
		{\tt exploglin} & $a + bx + c\log(x) + de^{x}$ & $a,b,c,d$ \\
		\hline
	\end{tabular}
	\caption{\footnotesize Table of nonlinear models used.}
	\label{tab:models}
\end{table}

For each dataset (one measurement) each model was fitted with the following procedure:
\begin{enumerate}
	\item For a given dataset (A-file), try to fit the given model via least squares procedure
	\item Compute metrics -- rMSE, MASE  
\end{enumerate}

In the first process we tried to remove the outliers from the initial fit. We tried removing 1 to 5 outliers (iteratively or at once) however, it did not lead to any significant improvement. Usually, the initial parts of the washout curve that were not fitted well were omitted leaving almost no difference to the terminal part in a refit curve.  
We implied that the level of 2.5\% and 5\%  is significant for medical specialist. When we follow the nitrogen curve in time beyond the 2.5\% level of concentration, it can be seen that the concentrations peaks can be interlaced with a nearly horizontal line. It is difficult for all models to fit properly such slowly decreasing end. That is why we also measured the quality of fit to a level where something is "still happening" (the curve does not decrease so slowly) -- up to 5\%.  
The rMSE results
can be seen in Tables \ref{tab:rmseall25} and \ref{tab:rmseall5}.

From the perspective of rMSE measure the model {\tt loglin} is the winner. The rMSE penalizes heavily the large misfits. If we take a look at the {\tt loglin} curve it can fit the initial part of the washout curve pretty well. All other models are penalized, except for the model {\tt exploglin}. It sometimes seems to be better, however, the coefficient in exponential member of the formula (d) is usually an extremely tiny number ($\sim 10^{-10}$). That is why this model is usually the same as \texttt{loglin}. From the perspective of Occam's principle further  we consider only the {\tt loglin} model. 

The curve with the best rMSE fit does not have to be necessary the best for the sake of prediction of the washout curve behaviour. It can be seen from the  MASE Tables \ref{tab:maseall25} and \ref{tab:maseall5}. 
If we compare the curves to the naive predictor, then we see that the {\tt exppow} model does often better than \texttt{loglin}. The model {\tt exp} model which is heavily used in describing the nitrogen washout curve in literature however is not so accurate. 

From the mentioned tables we can get the idea how the washout curve behaves. Model {\tt loglin} fits the data best. However, the tail of this model curve usually tends to grow up, we will see that later. This is not a plausible behaviour. Anyway, from the whole viewpoint this curve models the whole curve the best. 

We took an experiment and for all the curves tried to model only the first third of data. This way we could show that other models 
({\tt exp}, {\tt pow},  {\tt exppow}) are doing much better that the model {\tt loglin}. This brought us to idea that maybe the problem is in too shallow descend of the end of the washout curve that cannot be modeled well by any of used model curves.

When data sets were shortened up to the point where the nitrogen concentration decreases below 5\% of its initial concentrations, the model {\tt exppow} works much better on this initial phase. And its fitting  error improved. Nevertheless, the best fitting model is still  {\tt loglin}. We therefore have some candidates for interval fitting models -- the ones that have best rMSE and MASE at the same time. 
We omit the model {\tt exploglin}, since it is too complicated. We exclude the model {\tt log} since it is contained in {\tt loglin} and does not have better results than {\tt loglin}. We also cast out models {\tt explin} and {\tt explog} due to a large error rate. We have four remaining candidates -- {\tt exp}, {\tt pow}, {\tt exppow}, {\tt loglin} -- that we further use.   

 None of the checked model curves was able to nicely fit the data from the 5\% to 2.5\%. The level of 5\% seems to be the nice level that still enables possible plausible fitting with one of the classical models. This could also be an important fact for current discussions about advantages of LCI5 over LCI2.5. 
However, we must be careful not to reach the conclusions too quickly, because the part of the washout curve between 5\% and 2.5\% can possibly contain some important information about the quality of patient airways.  


\subsection{Interval models -- least squares}



\begin{figure}[h]
	\centering
	\includegraphics[width=12cm]{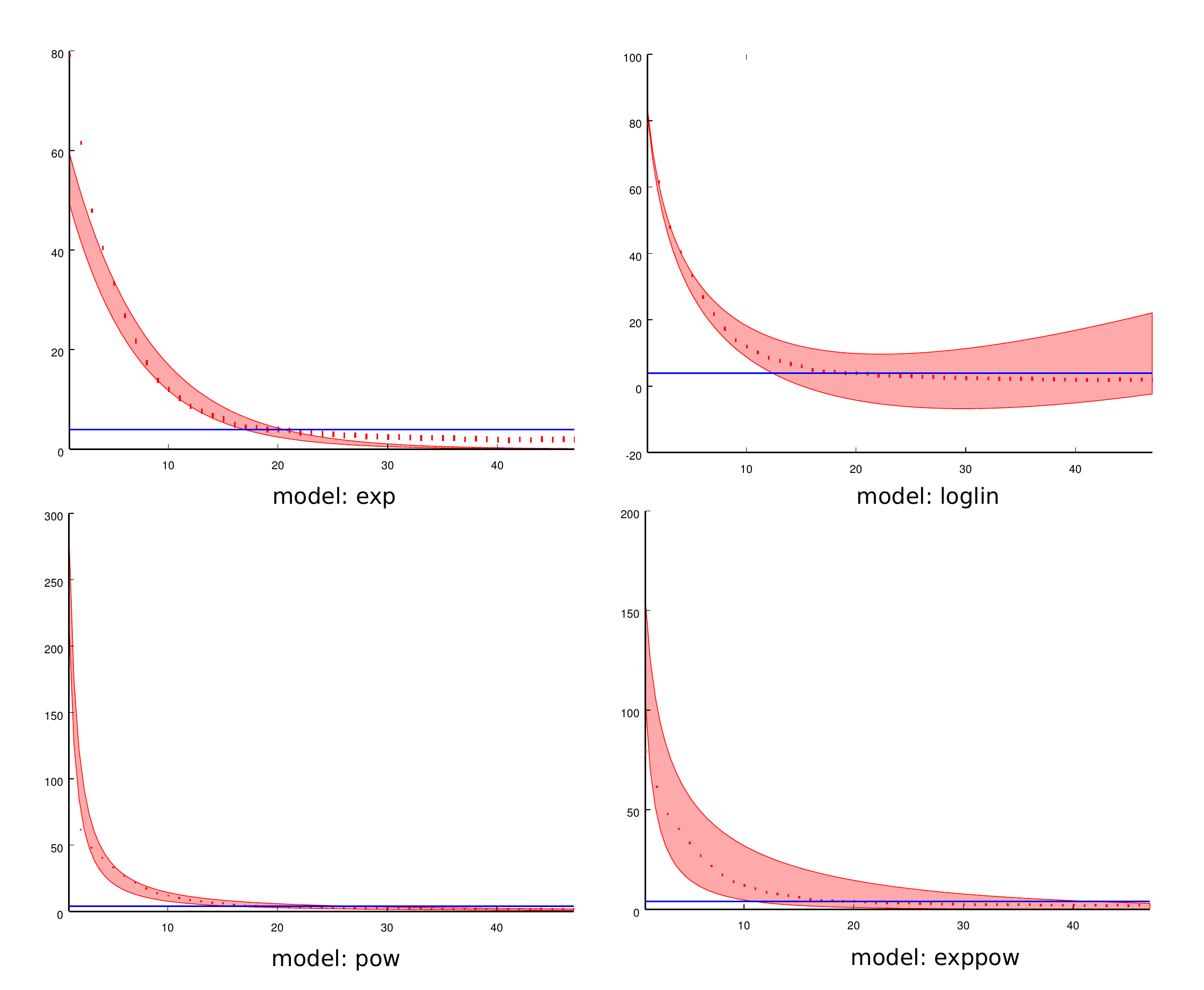}
	\caption{\footnotesize Interval curves fitting the real data with real measurement errors. The small red rectangles represent the interval data. The blue line represent the level of 5\% of initial nitrogen concentration. Notice that y-scale of each graph is different. Interval least squares fitting curves (interval washout models) are depicted in pink.}
	\label{fig:failfit}
\end{figure}

We took the four candidates on fitting curves -- {\tt exp}, {\tt pow}, {\tt exppow}, {\tt loglin} -- and provided the interval fitting. 
Each fitting can be transformed to solving an interval linear system of equations. The process is thoroughly described in \cite{cerny2013}. Unfortunately, the results were not encouraging. Due to the errors of sensors the interval data are consisted of intervals with large widths. That is why the resulting interval washout curve models are too thick. Another reason for such overestimation might be that solving an interval linear system exactly is a hard task (NP-hard in the language of computational complexity) therefore we usually use only approximative methods and they might provide some verified overestimations. 
Shapes typical for each interval washout model are depicted in Figure \ref{fig:failfit}. No curve was completely able to fit the data nicely. The {\tt exp} function misses the initial and final part of the washout data. The {\tt pow} model misses the initial part. 
The {\tt exppow} model is usually too wide, however, contains the data inside the interval curve. The {\tt loglin} model usually tends to widen in time, ruining any possibility of prediction. 

\subsection{Hypothetical sensors}

We showed that problem of the least squares fitting lies  within precision of current sensors 
(0.3\% for $O_2$ sensor and 5\% for $CO_2$ sensor of Exhalyzer D machine) and also possibly within the methods for solving interval systems of equations.  One might claim that the main flaw lies in the methods for solving interval systems and their overestimation. To shed more light on this, let us assume we have the sensors with better accuracy by one order i.e, 0.03\% for $O_2$ sensor and 0.5\% for $CO_2$ sensor.

\begin{figure}[h]
	\centering
	\includegraphics[width=12cm]{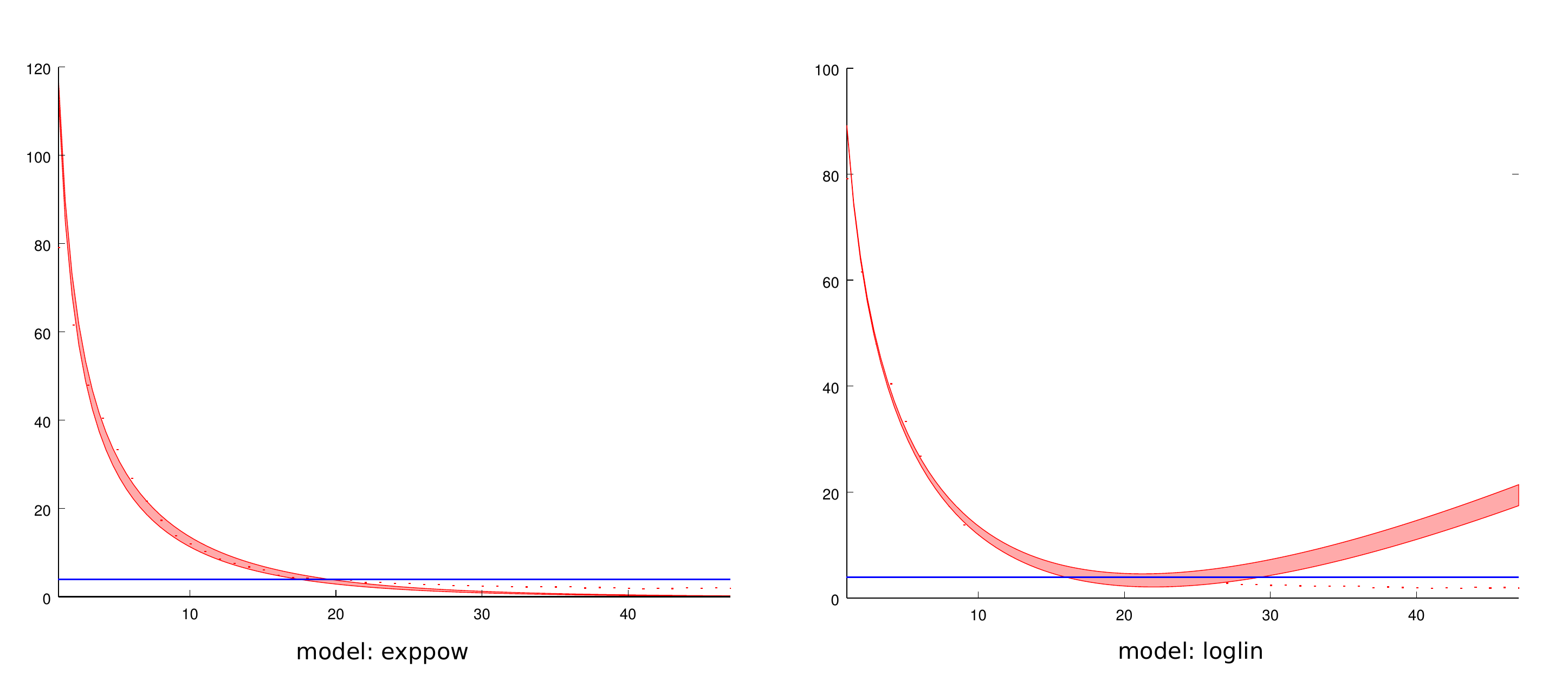}
	\caption{\footnotesize Interval curves fitting the real data with hypothetical measurement errors. The small red rectangles represent the interval data. The blue line represent the level of 5\% of initial nitrogen concentration. See the variable y-scale of each graph. Interval least squares fitting curves are depicted in pink.}
	\label{fig:okfit}
\end{figure}

Let us repeat the same procedure as in Figure \ref{fig:failfit}, this time for the hypothetical sensors. The surprising results are displayed in Figure \ref{fig:okfit}. We checked all the four mentioned models manually by visual evaluation.  We omitted the model {\tt pow}, because it gave poor fitting results in the initial parts. We also omitted the model {\tt exp}. Although, it gave very narrow curves it resulted in really poor fit. We checked the two remaining models -- {\tt exppow} and {\tt loglin}. 
The problems with {\tt loglin} still persist. Even for narrow intervals the curve tends to rise at its end. This gives us the winning description model -- {\tt exppow}. 
If we take a look at Figure \ref{fig:okfit}, we see that the behaviour of {\tt exppow} model does not fit the data well under the blue line (5\% concentration level). However, till the line it behaves well.  
We further check its properties in the next section.

\subsection{Prediction}

As it was said the level of nitrogen concentration where we stop the measurement is 2.5\% or 5\%. This boundary is set historically. For small uncooperative infants it might be difficult to prevent leaks and maintain calm and regular breathing for longer time. Sometimes the measurement must be aborted. In order to not waste the so far good measurement we can try to predict the successive behaviour of the washout curve. Using the previously developed interval washout models  we focus on determination of the terminal breath of a measurement. To remind the definition, for a given level of nitrogen concentration (20\%, 10\%, 5\% or 2.5\%),

the \emph{terminal breath} for this concentration is defined to be the first one of the three consecutive breaths with concentration below the respective level.       
 
\begin{figure}[h]
	\centering
	\includegraphics[width=10cm]{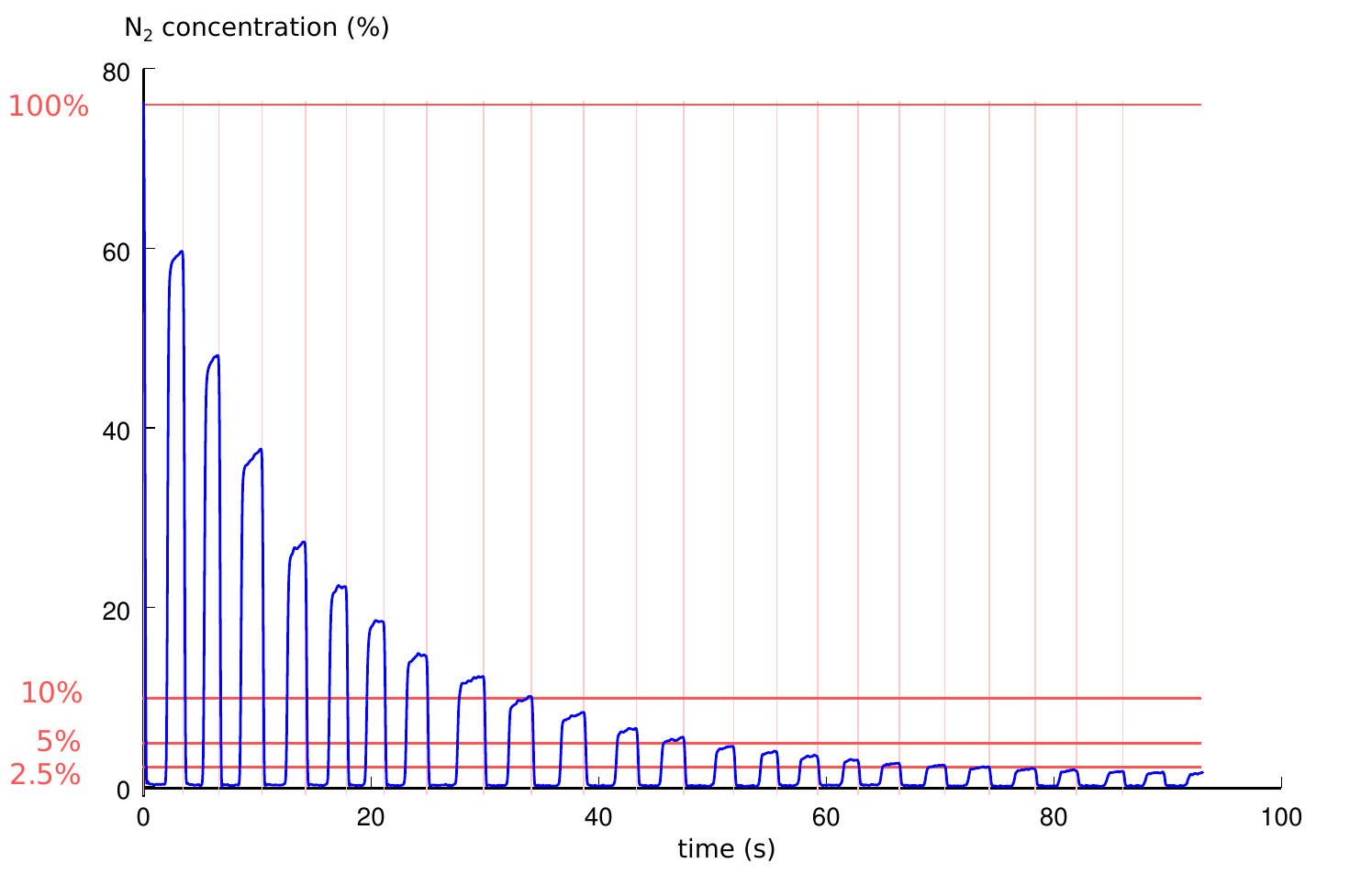}
	\caption{\footnotesize Concentrations of nitrogen in \% of the initial nitrogen concentration.}
	\label{fig:lci10}
\end{figure}

As discussed earlier, we limited our prediction to the part of the washout curve between 10\% and 5\% as depicted in Figure \ref{fig:lci10}. The goal was to predict the interval containing the terminal breath at 5\% level and compare it with the real terminal breath at the corresponding level. For the  prediction we used both the real and hypothetical sensors, the result are in Tables \ref{tab:realprediction}(real sensors) and \ref{tab:hypprediction} (hypothetical sensors). 

The case of real sensors is provided just for illustration, the resulting intervals predicting terminal breaths are too large. In the case of hypothetical sensors, the prediction is not generally bad. 
However, in some cases the prediction is completely wrong. We suppose that none of the tested models is completely suitable for absolutely correct prediction.
Nevertheless, the quality of prediction brings us to the very important question we tackle more in the following subsection. 

\subsection{An alternative clinical index?}
The prediction of washout curve in current software (Spiroware) is of poor quality. We could see that the prediction using verified interval regression is also not too trustworthy. The problem lies in an  unsatisfactory model of the nitrogen washout process.
We discussed many washout curve models, however none of them was plausible enough (for the purpose of prediction). Before one starts a hunt for better models, it needs to be specified, why exactly we need predictions and models of washout process.
One reason has been documented previously on an example of an interrupted measurement because of patient’s weak cooperation. 
Indeed, the possibility to predict washout process would be of a great clinical value. Unfortunately, our results indicate, that predictions are not possible within the currently used approach to washout data analysis. 

 Let us say we want to predict LCI from an incomplete measurement. To derive the LCI, the FRC is also needed. For FRC derivation we need to compute $V_{out}$ (as an integration of flow), therefore we need to know the missing flow data whose prediction is nearly impossible (too jagged shape of the flow curve). In conclusion, even if we had a good prediction, there is no way to compute 
 meaningful LCI with this prediction.

\begin{figure}[h]
	\centering
	\includegraphics[width=10cm]{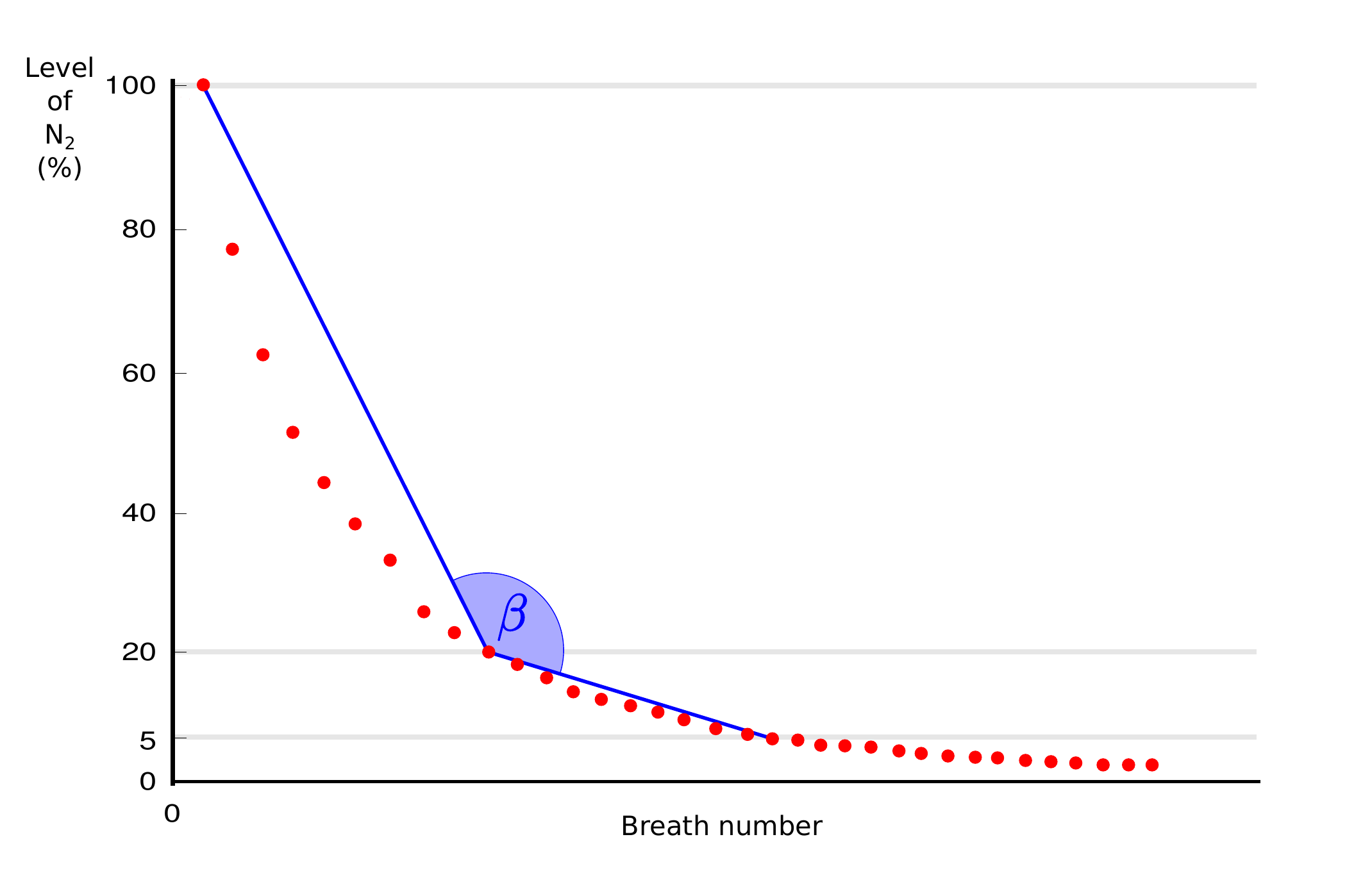}
	\caption{\footnotesize An example of an alternative hypothetical clinical index. The angle between two lines -- one going through the initial 100\% concentration of nitrogen and 20\%, the second going through 20\% and 5\%. }
	\label{fig:newindex}
\end{figure}

With that a new question arises --   can LCI be replaced by another index describing ventilation inhomogeneity and being more suitable to be predicted (and also robust enough to overcome some inaccuracy of prediction)? 
Much more suitable might be some form of clinical index that is based on the curvature of the washout curve. It would also permit to omit the computation of volume of air/nitrogen. During our early regression tests it seemed that for healthy persons the model \texttt{exppow} works better and for patients with cystic fibrosis the model \texttt{loglin} works better. We wanted to derive a new clinical index as a ratio of quality of the fit of these two methods. That is why all the tables contain the rightmost column "rat".  Another option would be an index $\beta$ depicted in Figure \ref{fig:newindex}. It is the angle of the two lines -- 
first going through the initial concentration and 20\% of concentration, the second going through 20\% concentration and 5\% of concentration.  However, these two indices remain hypothetical so far since the relation between them and lung properties is a subject of further clinical study.  

\section{Conclusions}

We summarize the results in the form of the following list:

\begin{itemize}
	\item We were able to significantly speed up the interval least squares procedure for certain specialized cases (e.g., the output data from MBW).
	\item An example of handling of uncertainties algebraically was shown.
	\item We demonstrated that the models that are usually used in literature for description of the behaviour of the nitrogen washout process are not plausible.
	\item We showed that if we consider the classical fitting models, the best model (but still not ideal) for the washout curve description is {\tt exppow}.
	\item Fitting the data with classical models up to 5\% is much more achievable than the attempts to fit the data up to 2.5\%.
	\item The current accuracy of Exhalyzer D sensors is insufficient for interval data estimation and making reasonable predictions.
	\item If we had sensors with better accuracy just by one order the verified fitting would work.
	\item It is impossible to predict the future value of LCI based on interrupted measurement due to properties of LCI.
	\item The possibility of new clinical indices was discussed.
\end{itemize} 

In our work numerous ways of future research emerged -- finding better models of the washout process, combination of the top-down and bottom-up approach in washout modeling, search for new clinical indices that will enable better prediction (our newly proposed indices are currently subjects of further clinical study). It would be also interesting to combine the
algebraic approach to uncertainty with the statistical one.

\bibliographystyle{plain}
\bibliography{literatura}

\begin{thebibliography}{10}

\bibitem{cerny2013}
Michal {\v{C}}ern{\'y}, Jarom{\'\i}r Antoch, and Milan Hlad{\'\i}k.
\newblock On the possibilistic approach to linear regression models involving
  uncertain, indeterminate or interval data.
\newblock {\em Information Sciences}, 244:26--47, 2013.

\bibitem{cotes2009lung}
John~E Cotes, David~J Chinn, and Martin~R Miller.
\newblock {\em Lung function: physiology, measurement and application in
  medicine}.
\newblock John Wiley \& Sons, 2009.

\bibitem{david2003clinical}
Yadin David, Wolf~W Von~Maltzahn, Michael~R Neuman, and Joseph~D Bronzino.
\newblock {\em Clinical engineering}.
\newblock CRC Press, 2003.

\bibitem{davies2008}
Jane~C Davies, Steve Cunningham, Eric~WFW Alton, and JA~Innes.
\newblock Lung clearance index in cf: a sensitive marker of lung disease
  severity.
\newblock {\em Thorax}, 63(2):96--97, 2008.

\bibitem{de2004new}
Francisco de~AT de~Carvalho, Eufrasio de A~Lima Neto, and Camilo~P Tenorio.
\newblock A new method to fit a linear regression model for interval-valued
  data.
\newblock In {\em Annual Conference on Artificial Intelligence}, pages
  295--306. Springer, 2004.

\bibitem{green2011}
Kent Green, Frederik~F Buchvald, June~Kehlet Marthin, Birgitte Hanel, Per~M
  Gustafsson, and Kim~Gjerum Nielsen.
\newblock Ventilation inhomogeneity in children with primary ciliary
  dyskinesia.
\newblock {\em Thorax}, pages thoraxjnl--2011, 2011.

\bibitem{gustafsson2007peripheral}
Per~M Gustafsson.
\newblock Peripheral airway involvement in cf and asthma compared by inert gas
  washout.
\newblock {\em Pediatric pulmonology}, 42(2):168--176, 2007.

\bibitem{heimlich2016gnu}
O~Heimlich.
\newblock Gnu octave interval package. version 1.4. 1, 2016.

\bibitem{Hla2014b}
Milan Hlad\'{\i}k.
\newblock New operator and method for solving real preconditioned interval
  linear equations.
\newblock {\em SIAM J. Numer. Anal.}, 52(1):194--206, 2014.

\bibitem{hladik2012}
Milan Hlad{\'\i}k and Michal {\v{C}}ern{\'y}.
\newblock Interval regression by tolerance analysis approach.
\newblock {\em Fuzzy Sets and Systems}, 193:85--107, 2012.

\bibitem{horacek:oils}
Jaroslav Hor{\'a}{\v c}ek and Milan Hlad{\'\i}k.
\newblock Computing enclosures of overdetermined interval linear systems.
\newblock {\em Reliable Computing}, 19:143, 2013.

\bibitem{Breathends}
Jaroslav Hor\'{a}{\v{c}}ek, V\'{a}clav Kouck\'{y}, and Milan Hlad\'{\i}k.
\newblock New insight into automated breath detection.
\newblock {\em In preparation.}

\bibitem{JauKie2001}
Luc Jaulin, Michel Kieffer, Olivier Didrit, and {\'E}ric Walter.
\newblock {\em Applied interval analysis. {With} examples in parameter and
  state estimation, robust control and robotics}.
\newblock Springer, London, 2001.

\bibitem{jensen2013standard}
Renee Jensen, Kent Green, Per Gustafsson, Philipp Latzin, Jessica Pittman,
  Felix Ratjen, Paul Robinson, Florian Singer, Sanja Stanojevic, and Sophie
  Yammine.
\newblock Standard operating procedure: multiple breath nitrogen washout.
\newblock {\em EcoMedics AG, Duernten, Switzerland}, 2013.

\bibitem{kearfott1996interval}
R~Baker Kearfott.
\newblock Interval computations: Introduction, uses, and resources.
\newblock {\em Euromath Bulletin}, 2(1):95--112, 1996.

\bibitem{macleod2009}
Kenneth~A Macleod, Alex~R Horsley, Nicholas~J Bell, Andrew~P Greening,
  J~Alastair Innes, and Steve Cunningham.
\newblock Ventilation heterogeneity in children with well controlled asthma
  with normal spirometry indicates residual airways disease.
\newblock {\em Thorax}, 64(1):33--37, 2009.

\bibitem{moore:introduction}
R.E. Moore, R.B. Kearfott, and M.J. Cloud.
\newblock {\em Introduction to interval analysis}.
\newblock Society for Industrial Mathematics, 2009.

\bibitem{neumaier1986linear}
Arnold Neumaier.
\newblock Linear interval equations.
\newblock In {\em Interval Mathematics 1985}, pages 109--120. Springer, 1986.

\bibitem{neumaier1990interval}
Arnold Neumaier.
\newblock {\em Interval Methods for Systems of Equations}.
\newblock Cambridge University Press, Cambridge, 1990.

\bibitem{robinson2012consensus}
Paul Robinson, Philipp Latzin, Sylvia Verbanck, Graham~L Hall, Alexander
  Horsley, Monika Gappa, Cindy Thamrin, Hubertus~GM Arets, Paul Aurora,
  S~Fuchs, et~al.
\newblock Consensus statement for inert gas washout measurement using multiple
  and single breath tests.
\newblock {\em European Respiratory Journal}, pages erj00697--2012, 2012.

\bibitem{rossing1967mathematical}
Robert~G Rossing, M~Bryan Danford, Earl~L Bell, and Raul Garcia.
\newblock Mathematical models for the analysis of the nitrogen washout curve.
\newblock Technical report, DTIC Document, 1967.

\bibitem{curvehandbook}
Vera Sit, Melanie Poulin-Costello, and Wendy Bergerud.
\newblock {\em Catalogue of curves for curve fitting}.
\newblock Forest Science Research Branch, Ministry of Forests, 1994.

\bibitem{tanaka1998interval}
Hideo Tanaka and Haekwan Lee.
\newblock Interval regression analysis by quadratic programming approach.
\newblock {\em IEEE Transactions on Fuzzy Systems}, 6(4):473--481, 1998.

\bibitem{tawhai2001multibreath}
Merryn~Howatson Tawhai and Peter~J Hunter.
\newblock Multibreath washout analysis: modelling the influence of conducting
  airway asymmetry.
\newblock {\em Respiration physiology}, 127(2-3):249--258, 2001.

\end{thebibliography}

\begin{table}[h]
	\centering
	\footnotesize
	\renewcommand\arraystretch{1.3} 
	\begin{tabular}{cccccccccccc}
		\hline
	no. & \texttt{exp} &\texttt{explin} &\texttt{pow} &\texttt{exppow} &\texttt{log} &\texttt{loglin} &\texttt{explin} &\texttt{explog} & \texttt{exploglin} & \texttt{rat} & H/CF\\[0.4ex]
		\hline
		1& 7.15 & 23.39 & 28.44 & 6.33 & 4.29 & \textbf{1.81} & 23.39 & 22.54 & 23.36 &  3.50 & H\\
		2&7.98 & 22.22 & 24.45 & 6.59 & 4.64 & \textbf{1.44} & 22.22 & 21.56 & 22.23 &  4.59& H \\
		3&10.67 & 11.84 & 11.30 & 8.11 & 6.26 & \textbf{1.16} & 11.84 & 5.70 & 0.97 &  6.98& H \\
		4&10.07 & 11.67 & 11.11 & 5.44 & 6.32 & \textbf{1.24} & 11.67 & 5.75 & 0.96 &  4.39 & H\\
		5&10.80 & 11.82 & 8.80 & 5.94 & 6.63 & \textbf{1.58} & 11.82 & 6.12 & 1.32 &  3.77& H \\
		6&8.86 & 10.85 & 16.38 & 5.79 & 4.82 & \textbf{1.24} & 10.85 & 4.68 & 0.77 &  4.68 & H\\
		7&10.71 & 19.15 & 5.91 & 9.44 & 5.71 & \textbf{1.09} & 19.15 & 19.15 & 19.15 &  8.64 & H\\
		8&7.00 & 10.40 & 21.37 & 4.63 & 4.25 & \textbf{1.18} & 10.40 & 3.83 & 1.18 &  3.91 & H\\
		9&7.10 & 10.60 & 20.71 & 3.81 & 4.43 & \textbf{1.27} & 10.60 & 4.05 & 1.26 &  2.99 & H\\
		10&3.44 & 23.75 & 31.45 & 2.35 & 2.67 & \textbf{2.18} & 23.75 & 23.75 & 23.75 &  1.08 & H\\
		11&4.27 & 25.36 & 35.95 & 3.04 & 3.03 & \textbf{2.16} & 25.36 & 25.36 & 25.36 &  1.41 & H\\
		12&3.61 & 24.58 & 33.17 & 2.14 & 2.58 & \textbf{1.84} & 24.58 & 24.58 & 24.58 &  1.16 & H\\
		13&2.74 & 26.09 & 36.08 & \textbf{1.13} & 2.29 & 1.87 & 26.09 & 26.09 & 26.09 &  0.60 & H\\
		14&5.30 & 10.11 & 24.22 & 2.44 & 3.82 & \textbf{1.46} & 10.11 & 3.40 & 1.45 &  1.67 & H\\
		15&10.30 & 16.14 & 6.34 & 7.50 & 5.96 & \textbf{1.97} & 16.14 & 16.14 & 16.14 &  3.81 & H\\
		\hline
		1& 8.01 & 11.17 & 13.76 & 2.58 & 5.98 & \textbf{1.34} & 11.17 & 5.16 & 1.16 &  1.93 & CF\\
		2&9.54 & 12.69 & \textbf{2.97} & 4.05 & 6.23 & 3.26 & 12.69 & 12.45 & 12.45 &  1.24 & CF\\
		3&10.32 & 33.33 & \textbf{2.71} & 7.47 & 6.46 & 2.91 & 33.33 & 13.55 & 33.33 &  2.57& CF \\
		4&9.08 & 11.69 & 11.43 & 3.85 & 7.00 & \textbf{1.32} & 11.69 & 5.78 & 0.81 &  2.93 & CF\\
		5&10.92 & 18.09 & 8.15 & 7.76 & 6.17 & \textbf{2.20} & 18.09 & 18.09 & 18.09 &  3.52 & CF\\
		6&8.31 & 14.40 & 5.36 & 3.38 & 4.69 & \textbf{1.90} & 14.40 & 14.40 & 14.40 &  1.78 & CF\\
		7&9.21 & 11.61 & 10.54 & 3.90 & 6.74 & \textbf{1.27} & 11.61 & 5.67 & 0.79 &  3.08 & CF\\
		8&7.55 & 11.04 & 22.42 & 5.86 & 4.89 & \textbf{1.66} & 11.04 & 4.35 & 1.66 &  3.53 & CF\\
		9&9.85 & 11.69 & 16.16 & 7.68 & 5.64 & \textbf{1.00} & 11.69 & 17.42 & 10.94 &  7.68 & CF\\
		10&7.38 & 10.78 & 23.83 & 5.41 & 4.53 & \textbf{1.03} & 10.78 & 4.07 & 1.03 &  5.24 & CF\\
		11&5.30 & 10.13 & 22.82 & 3.24 & 4.14 & \textbf{1.35} & 10.13 & 3.49 & 1.35 &  2.40 & CF\\
		12&7.40 & 10.88 & 19.37 & 4.57 & 5.04 & \textbf{1.14} & 10.88 & 4.43 & 1.11 &  4.02 & CF\\
		\hline		
	\end{tabular}
	\caption{\footnotesize rMSE for fitting up to 2.5\% of initial nitrogen concentration. A healthy person is marked with H, a patient with cystic fibrosis is marked with CF, \texttt{rat} shows the quality of fit of \texttt{exppow} divided by \texttt{loglin}.}
	\label{tab:rmseall25}
\end{table}

\begin{table}[h]
	\centering
	\footnotesize
	\renewcommand\arraystretch{1.3} 
	\begin{tabular}{cccccccccccc}
		\hline
	no. & \texttt{exp} &\texttt{explin} &\texttt{pow} &\texttt{exppow} &\texttt{log} &\texttt{loglin} &\texttt{explin} &\texttt{explog} & \texttt{exploglin} & \texttt{rat} & H/CF\\[0.4ex]
		\hline
		1&1.99 & 7.80 & 20.35 & \textbf{0.92} & 2.01 & 1.85 & 7.80 & 1.72 & 1.69 &  0.50 & H\\
		2&3.46 & 8.45 & 18.63 & 1.84 & 2.42 & \textbf{1.70} & 8.45 & 1.96 & 1.59 &  1.08 & H\\
		3&4.51 & 8.83 & 14.25 & 2.22 & 3.57 & \textbf{1.04} & 8.83 & 2.51 & 0.98 &  2.13 & H\\
		4&5.67 & 9.27 & 11.05 & 1.83 & 4.21 & \textbf{0.48} & 9.27 & 3.18 & 0.46 &  3.77 & H\\
		5&6.75 & 9.88 & 10.43 & 2.42 & 4.74 & \textbf{0.60} & 9.88 & 3.74 & 0.52 &  4.05 & H\\
		6&4.95 & 8.95 & 14.95 & 1.96 & 3.02 & \textbf{0.93} & 8.95 & 2.44 & 0.90 &  2.11& H \\
		7&10.71 & 19.15 & 5.91 & 9.44 & 5.71 & \textbf{1.09} & 19.15 & 19.15 & 19.15 &  8.64& H \\
		8&4.06 & 8.66 & 17.30 & 1.88 & 2.66 & \textbf{1.35} & 8.66 & 2.14 & 1.29 &  1.39 & H\\
		9&4.96 & 9.14 & 15.59 & 2.18 & 3.12 & \textbf{1.52} & 9.14 & 2.65 & 1.50 &  1.44 & H\\
		10&2.32 & 25.03 & 22.83 & \textbf{1.53} & 2.18 & 2.16 & 25.03 & 26.78 & 25.01 &  0.71 & H\\
		11&1.96 & 7.67 & 22.55 & \textbf{0.78} & 2.02 & 2.02 & 7.67 & 1.93 & 1.91 &  0.38 & H\\
		12&2.17 & 7.20 & 20.63 & \textbf{1.07} & 1.63 & 1.63 & 7.20 & 1.53 & 1.51 &  0.66 & H\\
		13&1.84 & 7.69 & 21.54 & \textbf{0.63} & 1.61 & 1.58 & 7.69 & 22.16 & 10.80 &  0.40 & H\\
		14&4.03 & 8.71 & 16.24 & 1.77 & 2.75 & \textbf{1.68} & 8.71 & 2.36 & 1.67 &  1.05 & H\\
		15&7.31 & 9.88 & 11.26 & 2.70 & 4.09 & \textbf{0.86} & 9.88 & 3.65 & 0.83 &  3.16 & H\\
		\hline
		1&5.59 & 9.65 & 11.00 & \textbf{0.94} & 4.78 & 1.15 & 9.65 & 3.83 & 0.98 &  0.81 & CF \\
		2&8.52 & 10.41 & 8.64 & 3.58 & 4.82 & \textbf{0.75} & 10.41 & 4.32 & 0.64 &  4.79 & CF\\
		3&7.50 & 10.06 & 10.10 & 2.88 & 4.39 & \textbf{0.68} & 10.06 & 3.85 & 0.59 &  4.24 & CF\\
		4&4.18 & 8.15 & 9.97 & 1.22 & 4.75 & \textbf{0.57} & 8.15 & 2.93 & 0.52 &  2.14 & CF\\
		5&7.39 & 10.14 & 12.08 & 3.67 & 4.29 & \textbf{0.88} & 10.14 & 3.69 & 0.88 &  4.17 & CF\\
		6&8.29 & 19.53 & 6.22 & 2.67 & 4.22 & \textbf{1.17} & 19.53 & 19.53 & 19.53 &  2.28 & CF\\
		7&6.01 & 9.44 & 10.36 & 2.25 & 5.31 & \textbf{0.67} & 9.44 & 3.71 & 0.59 &  3.36 & CF\\
		8&3.66 & 8.66 & 18.49 & 2.69 & 3.05 & \textbf{1.93} & 8.66 & 2.32 & 1.79 &  1.39 & CF\\
		9&4.84 & 9.07 & 14.72 & 1.57 & 3.15 & \textbf{0.93} & 9.07 & 2.54 & 0.92 &  1.68 & CF\\
		10&3.08 & 8.26 & 16.64 & \textbf{0.74} & 2.22 & 1.19 & 8.26 & 1.76 & 1.14 &  0.63 & CF\\
		11&2.72 & 8.17 & 17.33 & \textbf{1.20} & 2.66 & 1.48 & 8.17 & 1.97 & 1.38 &  0.81 & CF\\
		12&3.73 & 8.45 & 15.46 & 1.89 & 3.10 & \textbf{1.32} & 8.45 & 2.22 & 1.24 &  1.43 & CF\\
		\hline
	\end{tabular}
	\caption{\footnotesize rMSE for fitting up to 5\% of initial nitrogen concentration. A healthy person is marked with H, a patient with cystic fibrosis is marked with CF, \texttt{rat} shows the quality of fit of \texttt{exppow} divided by \texttt{loglin}.}
	\label{tab:rmseall5}
\end{table}

\begin{table}[h]
	\centering
	\footnotesize
	\renewcommand\arraystretch{1.3} 
	\begin{tabular}{cccccccccccc}
		\hline
	no. & \texttt{exp} &\texttt{explin} &\texttt{pow} &\texttt{exppow} &\texttt{log} &\texttt{loglin} &\texttt{explin} &\texttt{explog} & \texttt{exploglin} & \texttt{rat} & H/CF\\[0.4ex]
		\hline
		1&1.70 & 7.03 & 3.87 & 1.26 & 1.90 & \textbf{0.59} & 7.03 & 6.44 & 6.98 &  2.13 & H\\
		2&1.84 & 6.42 & 3.24 & 1.20 & 2.02 & \textbf{0.46} & 6.42 & 5.98 & 6.42 &  2.62 & H\\
		3&1.51 & 2.75 & 1.13 & 0.94 & 1.62 & \textbf{0.30} & 2.75 & 1.48 & 0.25 &  3.09 & H\\
		4&1.28 & 2.56 & 0.98 & 0.62 & 1.54 & \textbf{0.33} & 2.56 & 1.42 & 0.25 &  1.90 & H\\
		5&1.47 & 2.77 & 0.86 & 0.67 & 1.72 & \textbf{0.44} & 2.77 & 1.61 & 0.38 &  1.55 & H\\
		6&1.82 & 3.48 & 2.25 & 1.25 & 1.74 & \textbf{0.37} & 3.48 & 1.71 & 0.24 &  3.40 & H\\
		7&3.22 & 7.50 & 1.50 & 1.84 & 3.12 & \textbf{0.62} & 7.50 & 7.50 & 7.50 &  2.97 & H\\
		8&1.37 & 3.34 & 2.60 & 0.79 & 1.53 & \textbf{0.33} & 3.34 & 1.40 & 0.33 &  2.38 & H\\
		9&1.30 & 3.37 & 2.44 & 0.60 & 1.57 & \textbf{0.31} & 3.37 & 1.46 & 0.28 &  1.95 & H\\
		10&1.15 & 10.75 & 5.97 & \textbf{0.74} & 1.58 & 1.07 & 10.75 & 10.75 & 10.75 &  0.69 & H\\
		11&1.33 & 10.64 & 6.32 & \textbf{0.88} & 1.70 & 1.03 & 10.64 & 10.64 & 10.64 &  0.86 & H\\
		12&1.02 & 9.76 & 5.73 & \textbf{0.71} & 1.35 & 0.91 & 9.76 & 9.76 & 9.76 &  0.78 & H\\
		13&0.73 & 11.22 & 6.56 & \textbf{0.42} & 1.26 & 0.94 & 11.22 & 11.22 & 11.22 &  0.45 & H\\
		14&0.91 & 2.95 & 2.77 & 0.42 & 1.21 & \textbf{0.32} & 2.95 & 1.09 & 0.32 &  1.32 & H\\
		15&2.90 & 6.21 & 1.30 & 1.41 & 3.44 & \textbf{1.21} & 6.21 & 6.21 & 6.21 &  1.17 & H\\
		\hline
		1&0.70 & 1.85 & 0.99 & 0.27 & 1.06 & \textbf{0.23} & 1.85 & 0.93 & 0.20 &  1.19 & CF\\
		2&3.32 & 6.69 & 1.33 & \textbf{1.07} & 4.56 & 2.66 & 6.69 & 6.42 & 6.42 &  0.40 & CF\\
		3&3.30 & 10.45 & \textbf{1.07} & 1.40 & 4.40 & 2.18 & 10.45 & 5.99 & 10.45 &  0.64 & CF\\
		4&0.66 & 1.49 & 0.72 & 0.30 & 0.94 & \textbf{0.18} & 1.49 & 0.79 & 0.12 &  1.65 & CF\\
		5&2.41 & 5.10 & 1.39 & 1.37 & 2.63 & \textbf{0.93} & 5.10 & 5.10 & 5.10 &  1.47 & CF\\
		6&3.56 & 9.78 & 1.42 & \textbf{1.19} & 3.92 & 1.63 & 9.78 & 9.78 & 9.78 &  0.73 & CF\\
		7&0.75 & 1.61 & 0.71 & 0.34 & 1.00 & \textbf{0.19} & 1.61 & 0.84 & 0.12 &  1.75 & CF\\
		8&1.28 & 3.00 & 2.34 & 0.81 & 1.46 & \textbf{0.39} & 3.00 & 1.31 & 0.39 &  2.07 & CF\\
		9&1.92 & 4.36 & 1.99 & 1.18 & 2.17 & \textbf{0.36} & 4.36 & 4.59 & 3.12 &  3.23 & CF\\
		10&1.33 & 3.35 & 2.73 & 0.85 & 1.58 & \textbf{0.27} & 3.35 & 1.45 & 0.26 &  3.21 & CF\\
		11&0.73 & 2.30 & 2.16 & 0.42 & 1.03 & \textbf{0.26} & 2.30 & 0.88 & 0.27 &  1.61 & CF\\
		12&1.01 & 2.51 & 1.79 & 0.57 & 1.28 & \textbf{0.26} & 2.51 & 1.14 & 0.23 &  2.21 & CF\\
		\hline		
	\end{tabular}
	\caption{\footnotesize MASE for fitting up to 2.5\% of initial nitrogen concentration. A healthy person is marked with H, a patient with cystic fibrosis is marked with CF, \texttt{rat} shows the quality of fit of \texttt{exppow} divided by \texttt{loglin}.}
	\label{tab:maseall25}
\end{table}

\begin{table}[h]
	\centering
	\footnotesize
	\renewcommand\arraystretch{1.3} 
	\begin{tabular}{cccccccccccc}
		\hline
		no. & \texttt{exp} &\texttt{explin} &\texttt{pow} &\texttt{exppow} &\texttt{log} &\texttt{loglin} &\texttt{explin} &\texttt{explog} & \texttt{exploglin} & \texttt{rat} & H/CF\\[0.4ex]
		\hline
		1&0.32 & 1.86 & 2.44 & \textbf{0.16} & 0.53 & 0.45 & 1.86 & 0.45 & 0.42 &  0.36 & H\\
		2&0.56 & 2.06 & 2.16 & \textbf{0.30} & 0.63 & 0.41 & 2.06 & 0.53 & 0.40 &  0.74 & H\\
		3&0.42 & 1.30 & 1.06 & 0.25 & 0.55 & \textbf{0.14} & 1.30 & 0.40 & 0.14 &  1.73 & H\\
		4&0.48 & 1.24 & 0.80 & 0.17 & 0.60 & \textbf{0.06} & 1.24 & 0.45 & 0.05 &  3.00 & H\\
		5&0.61 & 1.41 & 0.76 & 0.24 & 0.72 & \textbf{0.09} & 1.41 & 0.58 & 0.07 &  2.77 & H\\
		6&0.66 & 1.99 & 1.58 & 0.30 & 0.73 & \textbf{0.19} & 1.99 & 0.60 & 0.19 &  1.59 & H\\
		7&3.22 & 7.50 & 1.50 & 1.84 & 3.12 & \textbf{0.62} & 7.50 & 7.50 & 7.50 &  2.97 & H\\
		8&0.59 & 2.00 & 1.93 & 0.31 & 0.66 & \textbf{0.30} & 2.00 & 0.54 & 0.29 &  1.03 & H\\
		9&0.74 & 2.13 & 1.69 & 0.30 & 0.78 & \textbf{0.27} & 2.13 & 0.66 & 0.27 &  1.09 & H\\
		10&0.70 & 9.64 & 4.32 & \textbf{0.47} & 0.97 & 0.91 & 9.64 & 10.53 & 9.63 &  0.52 & H\\
		11&0.46 & 2.88 & 3.86 & \textbf{0.22} & 0.71 & 0.71 & 2.88 & 0.65 & 0.68 &  0.30 & H\\
		12&0.32 & 2.49 & 3.57 & \textbf{0.28} & 0.60 & 0.60 & 2.49 & 0.55 & 0.56 &  0.46 & H\\
		13&0.40 & 3.13 & 3.99 & \textbf{0.21} & 0.59 & 0.61 & 3.13 & 7.60 & 3.62 &  0.34 & H\\
		14&0.59 & 1.94 & 1.75 & \textbf{0.29} & 0.63 & 0.33 & 1.94 & 0.54 & 0.33 &  0.88 & H\\
		15&1.12 & 2.46 & 1.34 & 0.44 & 1.12 & \textbf{0.19} & 2.46 & 1.02 & 0.18 &  2.28 & H\\
		\hline
		1&0.42 & 1.14 & 0.74 & \textbf{0.09} & 0.58 & 0.13 & 1.14 & 0.47 & 0.12 &  0.71 & CF\\
		2&1.26 & 2.43 & 0.93 & 0.51 & 1.23 & \textbf{0.18} & 2.43 & 1.13 & 0.15 &  2.84 & CF\\
		3&1.00 & 2.14 & 1.00 & 0.38 & 1.02 & \textbf{0.15} & 2.14 & 0.91 & 0.13 &  2.48 & CF\\
		4&0.23 & 0.68 & 0.53 & 0.07 & 0.40 & \textbf{0.05} & 0.68 & 0.25 & 0.04 &  1.60 & CF\\
		5&1.04 & 2.21 & 1.23 & 0.51 & 1.02 & \textbf{0.18} & 2.21 & 0.87 & 0.17 &  2.89 & CF\\
		6&2.60 & 9.57 & 1.24 & \textbf{0.68} & 2.39 & 0.71 & 9.57 & 9.57 & 9.57 &  0.97 & CF\\
		7&0.40 & 0.96 & 0.62 & 0.16 & 0.56 & \textbf{0.06} & 0.96 & 0.40 & 0.05 &  2.53 & CF\\
		8&0.51 & 1.65 & 1.73 & \textbf{0.35} & 0.60 & 0.37 & 1.65 & 0.45 & 0.36 &  0.94 & CF\\
		9&0.58 & 1.74 & 1.45 & 0.22 & 0.65 & \textbf{0.15} & 1.74 & 0.53 & 0.15 &  1.47 & CF\\
		10&0.38 & 1.64 & 1.67 & \textbf{0.12} & 0.48 & 0.21 & 1.64 & 0.37 & 0.20 &  0.59 & CF\\
		11&0.31 & 1.36 & 1.51 & \textbf{0.14} & 0.47 & 0.24 & 1.36 & 0.35 & 0.23 &  0.60 & CF\\
		12&0.41 & 1.33 & 1.25 & \textbf{0.21} & 0.51 & \textbf{0.21} & 1.33 & 0.37 & 0.20 &  0.97 & CF\\
		\hline		
	\end{tabular}
	\caption{\footnotesize MASE for fitting up to 5\% of initial nitrogen concentration. A healthy person is marked with H, a patient with cystic fibrosis is marked with CF, \texttt{rat} shows the ratios of quality of fit of \texttt{exppow} divided by \texttt{loglin}.}
	\label{tab:maseall5}
\end{table}

\begin{table}[h]
	\centering
	\footnotesize
	\renewcommand\arraystretch{1.3} 
	\begin{tabular}{cccccccccccc}
		\hline
		no. & len & real &\texttt{exp} &\texttt{pow} &\texttt{exppow} & \texttt{loglin} & H/CF\\ [0.4ex]
		\hline
		1 & 49 & 23 &\textbf{[21, 23]} & [49, 49] & [19, 26]& [15, 25] & H\\
		2 & 39 & 23 &[19, 21] & [39, 39] & [18, 25] & [14, 24] & H\\
		3 & 25 & 14 &[12, 13] & [19, 25] & \textbf{[11, 16]} & [9, 25] & H\\
		4 & 23 & 13 &[11, 12] & [18, 23] & \textbf{[10, 15]} & [8, 23] & H\\
		5 & 25 & 14 &[11, 12] & [18, 22] & [10, 16] & [8, 25] & H\\
		6 & 35 & 21 &[17, 18] & [32, 35] & [15, 23] & [12, 35] & H\\
		7 & 51 & 51 &[20, 22] & [42, 51] & [18, 28] & [14, 51] & H\\
		8 & 32 & 22 &[18, 20] & [32, 32] & [16, 24] & [13, 32] & H\\
		9 & 32 & 22 &[18, 20] & [32, 32] & [17, 26] & [13, 32] & H\\
		10 & 51 & 40 &[33, 37] & [51, 51] & [30, 43] & [22, 37] & H\\
		11 & 51 & 35 &\textbf{[32, 35]} & [51, 51] & [29, 41] & [22, 37] & H\\
		12 & 50 & 34 &\textbf{[30, 34]} & [50, 50] & [28, 39] & [21, 37] & H\\
		13 & 51 & 37 &\textbf{[35, 38]} & [51, 51] & [32, 45] & [24, 41] & H\\
		14 & 36 & 21 &[18, 20] & [36, 36] & [17, 25] & [13, 36] & H\\
		15 & 63 & 26 &[18, 20] & [32, 44] & [17, 27] & [13, 63] & H\\
		\hline 		
		1 & 28 & 12 &[10, 11] & [15, 19] & [9, 15] & [7, 28] & CF \\
		2 & 98 & 24 &[15, 17] & [27, 36] & [14, 23] & [11, 98] & CF\\
		3 & 80 & 21 &[15, 17] & [26, 35] & [14, 23] & [11, 80] & CF\\
		4 & 20 & 8 &\textbf{[7, 8]} & [11, 13] & \textbf{[7, 11]} & [5, 20] & CF\\
		5 & 48 & 22 &[15, 17] & [28, 37] & [14, 22] & [11, 48] & CF\\
		6 & 115 & 61 &[35, 39] & [69, 111] & [33, 73] & [22, 115] & CF\\
		7 & 23 & 10 &[8, 9] & [11, 14] & \textbf{[7, 12]} & [6, 23] & CF\\
		8 & 32 & 18 &[14, 16] & [28, 32] & \textbf{[13, 18]} & [11, 18] & CF\\
		9 & 40 & 19 &[16, 17] & [30, 40] & [15, 22] & [11, 40] & CF\\
		10 & 44 & 19 &\textbf{[17, 19]} & [33, 44] & [16, 23] & [13, 44] & CF\\
		11 & 26 & 16 &[14, 15] & [26, 26] & \textbf{[13, 18]} & [10, 23] & CF\\
		12 & 31 & 15 &[13, 14] & [22, 28] & \textbf{[12, 17]} & [9, 31] & CF\\
		\hline

	\end{tabular}
	\caption{\footnotesize Prediction from 10\% to 5\% -- real sensors; the intervals are predictions of the terminal breath number by  various interval models, len -- number of total breaths in file, real -- number of real breath end at 5\% breath, H -- healthy, CF -- cystic fibrosis.  Prediction intervals $[\ul{a}, \ol{a}]$ containing the true value of breath end having $|\ol{a} - \ul{a}| \leq 5$ are depicted in boldface.}
	\label{tab:realprediction}
\end{table}

\begin{table}[h]
	\centering
	\footnotesize
	\renewcommand\arraystretch{1.3} 
	\begin{tabular}{ccccccccccc}
		\hline
		no. & len & real &\texttt{exp} &\texttt{pow} &\texttt{exppow} & \texttt{loglin} & H/CF\\ [0.4ex]
		\hline
		1 & 49 & 23 &[22, 22] & [49, 49] & \textbf{[22, 23]} & [18, 19] & H \\
		2 & 39 & 23 &[20, 20] & [39, 39] & [20, 21] & [17, 18] & H\\
		3 & 25 & 14 &[12, 12] & [22, 22] & [12, 13] & [11, 12] & H\\
		4 & 23 & 13 &[11, 11] & [20, 20] & [12, 12] & [11, 23] & H\\
		5 & 25 & 14 &[11, 11] & [19, 20] & [12, 12] & [12, 25] & H\\
		6 & 35 & 21 &[17, 18] & [35, 35] & [18, 19] & [16, 17] & H\\
		7 & 51 & 51 &[21, 21] & \textbf{[49, 51]} & [22, 23] & [19, 22] & H\\
		8 & 32 & 22 &[19, 19] & [32, 32] & [19, 20] & [16, 17] & H\\
		9 & 32 & 22 &[19, 19] & [32, 32] & [20, 21] & [17, 19] & H\\
		10 & 51 & 40 &[35, 35] & [51, 51] & [35, 36] & [27, 29] & H\\
		11 & 51 & 35 &[33, 34] & [51, 51] & \textbf{[34, 35]} & [27, 28] & H\\
		12 & 50 & 34 &[32, 32] & [50, 50] & [32, 33] & [26, 28] & H\\
		13 & 51 & 37 &\textbf{[36, 37]} & [51, 51] & \textbf{[37, 38]} & [30, 31] & H\\
		14 & 36 & 21 &[19, 19] & [36, 36] & \textbf{[20, 21]} & [17, 18] & H\\
		15 & 63 & 26 &[19, 19] & [37, 38] & [21, 22] & [22, 63] & H\\
		\hline 		
		1 & 28 & 12 &[11, 11] & [17, 17] & \textbf{[11, 12]} & [28, 28] & CF\\
		2 & 98 & 24 &[16, 16] & [31, 32] & [17, 18] & [98, 98] & CF\\
		3 & 80 & 21 &[16, 16] & [30, 30] & [17, 18] & [80, 80] & CF\\
		4 & 20 & 8 &\textbf{[8, 8]} & [12, 12] & \textbf{[8, 8]} & [20, 20] & CF\\
		5 & 48 & 22 &[16, 16] & [31, 32] & [17, 18] & [15, 18] & CF\\
		6 & 115 & 61 &[37, 37] & [85, 89] & [45, 49] & [115, 115] & CF\\
		7 & 23 & 10 &[8, 8] & [12, 13] & [9, 9] & [23, 23] & CF\\
		8 & 32 & 18 &[15, 15] & [32, 32] & [15, 16] & [13, 13] & CF\\
		9 & 40 & 19 &[16, 17] & [34, 35] & [17, 18] & [15, 17] & CF\\
		10 & 44 & 19 &[18, 18] & [38, 39] & \textbf{[19, 20]} & [16, 18] & CF\\
		11 & 26 & 16 &[15, 15] & [26, 26] & [15, 15] & [13, 14] & CF\\
		12 & 31 & 15 &[13, 13] & [24, 25] & [13, 14] & [12, 13] & CF\\ 		
		\hline 		
	\end{tabular}
	\caption{\footnotesize Prediction from 10\% to 5\% -- hypothetical sensors;  the intervals are predictions of the terminal breath number by  various interval models, len -- number of total breaths in file, real -- number of real breath end at 5\% breath, H -- healthy, CF -- cystic fibrosis.  Prediction intervals $[\ul{a}, \ol{a}]$ containing the true value of breath end having $|\ol{a} - \ul{a}| \leq 2$ are depicted in boldface.}
	\label{tab:hypprediction}
\end{table}

\end{document}